\documentclass[extra]{gji}
\usepackage{timet}
\usepackage{lineno}
\usepackage{graphicx}
\usepackage{amsmath,amsfonts,amssymb}
\usepackage{breqn}
\usepackage[normalem]{ulem}
\usepackage[dvipsinames]{xcolor}
\usepackage{longtable}
\usepackage{enumitem}

\usepackage[%
           colorlinks=True,%
           urlcolor=blue,%
           linkcolor=blue,%
           citecolor= blue,%
           pdfauthor={T. Schwaiger},%
           pdftitle={Relating force balances and flow length scales in geodynamo simulations},]{hyperref}
\usepackage{booktabs}
\usepackage{hyperref}

\DeclareMathOperator*{\argmax}{argmax}
\DeclareMathOperator*{\argmin}{argmin}

\definecolor{darkgreen}{rgb}{0,0.57,0.04}

\let\oldalign\align
\let\oldendalign\endalign
\renewenvironment{align}
  {\linenomathNonumbers\oldalign}
  {\oldendalign\endlinenomath}

\title[Relating force balances and flow length scales in geodynamo simulations]
  {Relating force balances and flow length scales in geodynamo simulations}
\author[T. Schwaiger et al.]
  {T. Schwaiger$^1$, T. Gastine$^1$ and J. Aubert$^1$ \\
  $^1$ Universit\'e de Paris, Institut de Physique du Globe de Paris, CNRS, F-75005 Paris, France.
  }

\date{Received \today; in original form \today}
\pagerange{\pageref{firstpage}--\pageref{lastpage}}

\begin{document}

\label{firstpage}

\maketitle

\begin{summary}
In fluid dynamics, the scaling behaviour of flow length scales is commonly used to infer the governing force balance of a system. The key to a successful approach is to measure length scales that are simultaneously representative of the energy contained in the solution (energetically relevant) and also indicative of the established force balance (dynamically relevant). In the case of numerical simulations of rotating convection and magneto-hydrodynamic dynamos in spherical shells, it has remained difficult to measure length scales that are both energetically and dynamically relevant, a situation that has led to conflicting interpretations, and sometimes misrepresentations of the underlying force balance. By analysing an extensive set of magnetic and non-magnetic models, we focus on two length scales that achieve both energetic and dynamical relevance. The first one is the peak of the poloidal kinetic energy spectrum, which we successfully compare to crossover points on spectral representations of the force balance.  In most dynamo models, this result confirms that the dominant length scale of the system is controlled by a previously proposed quasi-geostrophic (QG-) MAC (Magneto-Archimedean-Coriolis) balance. In non-magnetic convection models, the analysis generally favours a QG-CIA (Coriolis-Inertia-Archimedean) balance. Viscosity, which is typically a minor contributor to the force balance, does not control the dominant length scale at high convective supercriticalities in the non-magnetic case, and in the dynamo case, once the generated magnetic energy largely exceeds the kinetic energy. In dynamo models, we introduce a second energetically relevant length scale associated with the loss of axial invariance in the flow. We again relate this length scale to another crossover point in scale-dependent force balance diagrams, which marks the transition between large-scale geostrophy (the equilibrium of Coriolis and pressure forces) and small-scale magnetostrophy, where the Lorentz force overtakes the Coriolis force. Scaling analysis of these two energetically and dynamically relevant length scales suggests that the Earth's dynamo is controlled by a QG-MAC balance at a dominant scale of about $200 \, \mathrm{km}$, while magnetostrophic effects are deferred to scales smaller than $50 \, \mathrm{km}$.
\end{summary}


\section{Introduction}
Turbulent convective motions of liquid metal in the Earth's outer core initiate dynamo action that maintains the geomagnetic field. 
Based on the magneto-hydrodynamic theory of convection-driven dynamos, it is known that the fluid flow in the Earth's core is influenced by six different forces, namely Coriolis, pressure, buoyancy, Lorentz, inertial and viscous forces. Our main tool to study their relative importance and thus to know  which dynamical regime the Earth's core is in, are dimensionless numbers, which can be computed from the core's physical properties (see Table~\ref{table_parameter_values}). These can be interpreted as order of magnitude estimates of the ratio between two forces. Since the Earth is a rapidly-rotating system, the relative strengths of the individual forces with respect to the Coriolis force are considered. In the following, we adopt the thickness of the outer core $L$ as a characteristic flow length scale. The dimensionless numbers therefore only provide an estimate of the global-scale force balance. It can be expected that the force balance differs at smaller scales.

The ratio between viscous and Coriolis forces can be estimated with the help of the Ekman number
\begin{align}
E = \frac{\nu}{\Omega L^{2}} \sim \mathcal{O} \left( 10^{-15} \right),
\end{align}
while the relative amplitude of inertia compared to Coriolis forces is given by the Rossby number 
\begin{align}
Ro = \frac{U}{\Omega L} \sim \mathcal{O} \left( 10^{-6} \right).
\end{align}
In the above two definitions, $\Omega$ denotes the Earth's rotation rate, $U$ a characteristic flow velocity and $\nu$ the kinematic viscosity. The smallness of $E$ and $Ro$ allows us to conclude that in the Earth's core inertial and in particular viscous contributions are insignificant compared to rotational effects at large scales.
The strength of buoyancy relative to Coriolis forces can be estimated by defining the following buoyancy number
\begin{align}
Bu = \frac{\alpha \Theta g_{\mathrm{o}}}{\Omega U} \sim \mathcal{O} \left( 10^{-1} \right),
\end{align}
where $\alpha$ represents the thermal expansion coefficient, $\Theta$ a superadiabatic temperature perturbation and $g_{\mathrm{o}}$ the gravity at the core-mantle boundary. 
The ratio between Lorentz and Coriolis forces in planetary dynamos is often assessed by evaluating the traditional form of the Elsasser number, which is defined by
\begin{align}
\Lambda_{\mathrm{t}} = \frac{B^{2}}{\rho \mu \lambda \Omega} \sim \mathcal{O} \left( 10 \right),
\end{align}
where $B$ is the magnetic field strength, $\rho$ the fluid density, $\mu$ the magnetic permeability and $\lambda$ the magnetic diffusivity. This definition suggests a dominant role of the Lorentz force in governing the flow dynamics. However, it has been argued that the traditional Elsasser number is an unreliable measure of the force ratio due to some of the underlying assumptions possibly not being fully satisfied in turbulent dynamos \citep[see e.g.][]{soderlund_etal_2012,soderlund_etal_2015,dormy_2016}.
A more exact estimate can be obtained using the so-called dynamic Elsasser number \citep[e.g.][]{christensen_etal_1999,cardin_etal_2002,soderlund_etal_2012,soderlund_etal_2015}
\begin{align}
\Lambda_{d} = \frac{B^{2}}{\rho \mu \Omega U L} \sim \mathcal{O} \left( 10^{-2} \right).
\label{dyn_els}
\end{align}
In contrast to $\Lambda_{t}$, the dynamic Elsasser number indicates that the Lorentz force is two orders of magnitude smaller than the Coriolis force, suggesting that convection dynamics are rotationally-dominated. The final force that has not been considered so far is the one due to pressure. In the case of negligible inertia and viscosity, the pressure force compensates the part of the Coriolis force that is not balanced by buoyancy and Lorentz forces.
The hierarchy of the forces indicated by dimensionless numbers is different depending on which Elsasser number is used to estimate the Lorentz force. It is clear, however, that in either case viscosity and inertia represent the least important contributions, since they are far smaller than the other four forces. When considering the dynamic Elsasser number, buoyancy and Lorentz forces come about one or two orders of magnitude below the prevailing force equilibrium between pressure and Coriolis forces. The leading-order force balance in this case is therefore geostrophic \citep[e.g.][]{busse_1970}. The traditional Elsasser number, on the other hand, suggests a much stronger Lorentz force, so that the dominant force balance would be between pressure, Coriolis and Lorentz forces, which is termed magnetostrophic (MS) balance \citep[e.g.][]{roberts_1978}. As already pointed out above, the non-dimensional numbers only provide information about the force balance on the system scale. The true force balance in the Earth's core is likely more complicated due to the length-scale dependence of the forces \citep[e.g.][]{aurnou_and_king_2017,aubert_etal_2017}. Based on scaling analysis of $\Lambda_{d}$ (Eq. \ref{dyn_els}), which is described in detail in Section \ref{elsasser_scaling}, \citet{aurnou_and_king_2017} for instance argued that only the large-scale flow is predominantly geostrophic, while magnetostrophy occurs on smaller scales.

\begin{table*}
\caption{Typical estimates of physical properties of the Earth's outer core.}
\centering
\begin{tabular}{llll}
\hline
Symbol 			& Definition 								& Value & Reference \\ \hline
$\Omega$ 		& Rotation rate  							& $7.29 \times 10^{-5} \, \mathrm{s}^{-1}$ \\[1.1mm]
$L$				& Thickness of outer core					& $2.26 \times 10^{6} \, \mathrm{m}$  & \citet{dziewonski_and_anderson_1981} \\[1.1mm]
$\rho$			& Mean core density							& $1.1 \times 10^{4} \, \mathrm{kg} \, \mathrm{m}^{-3}$ & \citet{dziewonski_and_anderson_1981} \\[1.1mm]
$g_{o}$	& Gravity at core-mantle boundary			& $10.68 \, \mathrm{m} \, \mathrm{s}^{-2}$ & \citet{dziewonski_and_anderson_1981} \\[1.1mm]
$\mu$			& Magnetic permeability 						& $4 \pi \times 10^{-7} \, \mathrm{H} \, \mathrm{m}^{-1}$ \\[1.1mm]
$\nu$			& Kinematic viscosity					& $10^{-6} \, \mathrm{m}^{2} \, \mathrm{s}^{-1}$ & \citet{pozzo_etal_2013} \\[1.1mm]
$\alpha$			& Thermal expansion coefficient				& $10^{-5} \, \mathrm{K}^{-1} $ & \citet{gomi_etal_2013} \\[1.1mm]
$\Theta$		& Typical superadiabatic temperature perturbation		& $10^{-4} \, \mathrm{K}$ & \citet{jones_2015} \\[1.1mm]
$c_{p}$			& Specific heat capacity at constant pressure			& $850 \, \mathrm{J} \, \mathrm{kg}^{-1} \, \mathrm{K}^{-1}$ & \citet{stacey_1993} \\[1.1mm]
$k$			& Thermal conductivity						& $100 \, \mathrm{W} \, \mathrm{m}^{-1} \, \mathrm{K}^{-1} $ & \citet{pozzo_etal_2012} \\[1.1mm]
$\kappa$			& Thermal diffusivity						& $10^{-5} \, \mathrm{m}^{2} \, \mathrm{s}^{-1}$ & $\kappa = k / \rho c_{p} $ \\[1.1mm]
$\lambda$		& Magnetic diffusivity					& $0.7 \, \mathrm{m}^{2} \, \mathrm{s}^{-1}$ & \citet{pozzo_etal_2012} \\[1.1mm]
$U$				& Typical flow velocity						& $4 \times 10^{-4} \, \mathrm{m} \, \mathrm{s}^{-1}$  & \citet{finlay_and_amit_2011} \\[1.1mm]
$B$				& Typical magnetic field strength			& $4 \times 10^{-3} \, \mathrm{T}$ & \citet{gillet_etal_2010} \\[1.2mm] \hline
\end{tabular}
\label{table_parameter_values}
\end{table*}

The nature of the leading-order force balance has a significant effect on the convective pattern. A dominant geostrophic balance would result in convection being primarily organised into columnar eddies that are aligned with the rotation axis as a result of the so-called Proudman-Taylor theorem. In the case of magnetostrophy, on the other hand, the Lorentz force would be able to relax this rotational constraint, and thus break up the columnarity of the flow, resulting in larger scale flow. Core flow inversions based on observations of the secular variation of the Earth's magnetic field show consistency with leading-order geostrophy \citep[e.g.][]{pais_and_jault_2008, gillet_etal_2012, aubert_2020}. However, solutions to the core flow problem are non-unique, and the unknown contribution from unresolved length scales in the magnetic field and secular variation limits the spatial resolution of the inverted flows to spherical harmonic degrees well below $\ell = 14$ \citep[e.g.][]{hulot_etal_2015}. Therefore, smaller unresolvable scales could still be in a magnetostrophic state.

In addition to observations and theoretical considerations, numerical simulations are an important tool for our understanding of the geodynamo mechanism. Computational constraints make it currently impossible to simulate the extreme range of spatial and temporal scales present in the Earth's core \citep[e.g.][]{schaeffer_etal_2017}. 
Nonetheless, dynamo models can provide valuable insights into our planet's core dynamics if they operate in a relevant force balance regime, since this would allow a meaningful extrapolation of the numerical results to realistic parameters. Two different approaches have been used to assess the dynamical regime such simulations operate in. First, the comparison of the scaling behaviour of measured flow length scales to scaling laws, which allows one to infer the underlying force balance. Second, the explicit calculation of the magnitude of the individual forces.

The analysis of the scaling behaviour of convective length scales in geodynamo simulations has not resulted in an agreement with a scaling law based on a geophysically-relevant force balance (see Section \ref{scaling_laws} for details about the scaling laws). Instead, it has been shown that typical flow length scale measures roughly follow the viscous scaling \citep[e.g.][]{king_and_buffett_2013,oruba_and_dormy_2014}. This has led to the suggestion that convection in numerical dynamos is viscously-controlled and therefore not applicable to the Earth's core.

Explicit calculations of the root-mean-square (r.m.s.) strength of the forces in the dynamo models \citep[e.g.][]{wicht_and_christensen_2010,soderlund_etal_2012,yadav_etal_2016,aubert_etal_2017,aubert_2019,schwaiger_etal_2019} did not confirm a dominant role of viscosity. Instead, these studies showed that the force equilibrium in geodynamo simulations is typically composed of a zeroth-order balance between Coriolis and pressure forces, followed by a balance between buoyancy, Lorentz and ageostrophic Coriolis forces. This type of force balance has been referred to as quasi-geostrophic Magneto-Archimedean-Coriolis (QG-MAC) balance \citep{aubert_2019,schwaiger_etal_2019}. Inertia and viscosity were found to be second-order contributions, although the difference between them and the first-order forces proved to be rather small in models computed at moderate control parameters, in particular if viscous boundary layers are not excluded \citep{soderlund_etal_2012}. Recent high-resolution simulations \citep[e.g.][]{yadav_etal_2016,aubert_etal_2017,schaeffer_etal_2017,aubert_2019} in advanced parameter regimes showed that the relative importance of viscosity and inertia decreases in more realistic setups. The leading-order structure of the force balance, however, remains essentially unchanged. While these studies brought more insight to the discussion about the underlying physics in numerical dynamos, it has still remained an outstanding task to successfully relate force balances to convective flow length scales.

The contradiction between the results obtained by applying theoretical scalings to a flow length scale and the explicit force balance calculation is likely the result of two often overlooked questions: 
(\textit{i}) How do we estimate length scales that are sufficiently representative of the energy contained in the solution? 
(\textit{ii}) Is the length scale measure representative of the underlying force equilibria?

The recent introduction of a spectral analysis of the forces by \citet{aubert_etal_2017} has provided access to the length scale dependence of the force balance. This approach revealed that the zeroth-order force balance in geophysically-relevant numerical models is either geostrophic at all scales, or subdivided into large-scale geostrophy and small-scale magnetostrophy. Similarly, in the first-order MAC equilibrium the ageostrophic Coriolis force is predominantly balanced by buoyancy at large scales, and by the Lorentz force towards smaller scales. The respective transitions between large- and small-scale balances define triple points, at which three forces are of comparable magnitude. The associated length scales are referred to as cross-over length scales, which are by construction characteristic of the underlying physics. 
The goal of this paper is to relate them to energetically relevant flow length scales, as well as to theoretical scaling laws. To this end, we will analyse a series of dynamo models as well as non-magnetic rotating convection models for comparison. The study presented here is a follow-up to \citet{schwaiger_etal_2019}, where the force balance in the considered dynamo models was systematically analysed, and to \citet{aubert_etal_2017} and \citet{aubert_2019}, in which the force balance tools and the concept of cross-over length scales were introduced.

We outline the various scaling laws that have been proposed for relevant convective flow length scales in Section \ref{scaling_laws}. In Section \ref{methods}, we describe the numerical models and methods. The results of our study are presented and discussed in Sections \ref{results} and \ref{discussion}. Our conclusions are summarised in Section \ref{conclusions}. Note that throughout the manuscript, when we use the terms ``dominant'' or ``leading-order'' force balance, we refer to the zeroth-order force equilibrium.

\section{Theoretical scalings of the flow length scale}
\label{scaling_laws}
At this point we would like to recollect several theoretical scalings of relevant length scales that have been suggested for dynamos and non-magnetic rotating convection based on different assumptions regarding the governing force balance. The starting point for these scalings is the Navier-Stokes equation for thermally-driven Boussinesq convection in a rotating reference frame:
\begin{align}
\begin{split}
\underbrace{\rho \left( \frac{\partial \mathbf{u}}{\partial t} + \mathbf{u} \cdot \nabla \mathbf{u} \right)}_{ \text{\normalfont inertia}} + \underbrace{2 \rho \mathbf{\Omega} \times \mathbf{u}}_{ \text{\normalfont Coriolis}} = & \underbrace{-\nabla P}_{\text{\normalfont pressure}} + \underbrace{\rho \alpha T \mathbf{g}}_{\text{\normalfont buoyancy}} + \\ & + \underbrace{\mathbf{j} \times \mathbf{B}}_{\text{\normalfont Lorentz}} + \underbrace{\rho \nu \nabla^{2} \mathbf{u}}_{\text{\normalfont viscosity}},
\end{split}
\label{Navier_Stokes_dim}
\end{align}
where $\mathbf{u}$ is the velocity field, $\mathbf{\Omega}$ the rotation vector, $P$ the pressure, $T$ the temperature, $\mathbf{g}$ the gravitational  acceleration, $\mathbf{B}$ the magnetic field and $\mathbf{j} = 1/\mu \nabla \times \mathbf{B}$ the electric current density. The labels denote the corresponding forces.
\subsection{Leading-order geostrophy}
Since the Earth's core is rapidly rotating, the leading-order force balance is often assumed to be geostrophic \citep[e.g.][]{busse_1970,calkins_2018}, that is a balance between the Coriolis force and the pressure gradient:
\begin{align}
2 \rho \mathbf{\Omega} \times \mathbf{u} = - \nabla P.
\end{align}
Taking the curl yields
\begin{align}
\frac{\partial \mathbf{u}}{\partial z} = 0,
\label{Proudman_Taylor}
\end{align}
which is known as the Proudman-Taylor theorem. It constrains the fluid flow to the two dimensions perpendicular to the rotation axis. The only truly geostrophic flows in spherical shells are however axisymmetric, and as such cannot drive the geodynamo according to Elsasser's anti-dynamo theorem \citep{elsasser_1946}. Therefore, poloidal flow, which can only arise from deviations to geostrophy, is required in the Earth's core.
This non-geostrophic (ageostrophic) flow can be obtained from (\ref{Navier_Stokes_dim}) by taking its curl to remove the pressure gradient, which yields the vorticity equation
\begin{align}
\begin{split}
\frac{\partial \boldsymbol{\omega}}{\partial t} + \mathbf{u} \cdot \nabla \boldsymbol{\omega} - \boldsymbol{\omega} \cdot \nabla \mathbf{u} + 2 \Omega \frac{\partial \mathbf{u}}{\partial z} = & \nabla \times \left( \alpha T \mathbf{g} \right) + \\  + \frac{1}{\rho} \nabla \times \left( \mathbf{j} \times \mathbf{B} \right) + \nu \nabla^{2} \boldsymbol{\omega},
\end{split}
\label{vorticity_equation}
\end{align}
where $\boldsymbol{\omega} = \nabla \times \mathbf{u}$ is the vorticity.

Based on the assumption of triple force balances in (\ref{vorticity_equation}), three different scalings for the convective flow length scale have been suggested. All three triple balances have in common the thermal wind balance, that is the balance between ageostrophic Coriolis and buoyancy terms, termed AC in the following. It is supplemented by a third force term that is hypothesised to be responsible for breaking the Proudman-Taylor constraint. This third force could be viscosity, inertia or the Lorentz force forming the respective VAC, CIA and MAC force balance. By virtue of the small Ekman and Rossby numbers in the Earth's core, the Lorentz force is considered to be the geophysically most relevant candidate. Nevertheless, we will outline all three scalings below. Since the parameters in geodynamo simulations are far from Earth's conditions, the dynamics might be controlled by inertia or viscosity in numerical models. In addition, for the purpose of comparison our study includes a series of non-magnetic rotating convection simulations, for which obviously only scalings that do not involve the Lorentz force are relevant.

\subsubsection{VAC balance}
Assuming a balance between viscous, buoyancy and Coriolis terms, which is sometimes referred to as the VAC balance (VAC for Viscous-Archimedean-Coriolis), in (\ref{vorticity_equation}) and using $\mathcal{L}_{\perp}$ and $\mathcal{L}_{\parallel}$ to describe the integral length scales perpendicular and parallel to the rotation axis, respectively, yields the following order of magnitude estimates
\begin{align}
\frac{\nu \omega}{\mathcal{L}_{\perp}^{2}} \sim \frac{\alpha \Theta g_{\mathrm{o}}}{\mathcal{L}_{\perp}} \sim \frac{\Omega U}{\mathcal{L}_{\parallel}}.
\end{align}
The zeroth-order geostrophic balance leads to convection predominantly occuring in columnar vortices aligned with the rotation axis. Therefore, it can be assumed that $\mathcal{L}_{\parallel} \sim L$. Combining this with the assumption $\omega \sim U/\mathcal{L}_{\perp}$, the balance between Coriolis and viscous terms yields 
\begin{align}
\frac{\mathcal{L}_{\perp}}{L} \sim \left( \frac{\nu}{\Omega L^{2}} \right)^{1/3} = E^{1/3},
\label{viscous_scaling}
\end{align}
which corresponds to the flow length scale at the onset of convection \citep[e.g.][]{busse_1970,king_and_buffett_2013}.

\subsubsection{CIA balance}
The combination of large Reynolds ($Re = UL/\nu \sim \mathcal{O}  \left( 10^{9} \right)$) and low Rossby numbers in the Earth's core has resulted in some studies considering a turbulent quasi-geostrophic balance between Coriolis, inertia and buoyancy terms in (\ref{vorticity_equation}). The acronym CIA (CIA for Coriolis-Inertia-Archimedean) is commonly used to refer to this type of balance.
The amplitudes of the individual terms are given by
\begin{align}
\frac{\Omega U}{\mathcal{L}_{\parallel}} \sim \frac{U \omega}{\mathcal{L}_{\perp}} \sim \frac{\alpha \Theta g_{\mathrm{o}}}{\mathcal{L}_{\perp}}.
\end{align}
Combining the assumption $\mathcal{L}_{\parallel} \sim L$ with the balance between curled Coriolis and inertial forces leads to
\begin{align}
\frac{\mathcal{L}_{\perp}}{L} \sim \frac{\omega}{\Omega}.
\end{align}
Assuming again $\omega \sim U/\mathcal{L}_{\perp}$ yields the following prediction for the integral flow length scale $\mathcal{L}_{\perp}$
\begin{align}
\frac{\mathcal{L}_{\perp}}{L} \sim \left( \frac{U}{\Omega L} \right)^{1/2} \sim Ro^{1/2},
\label{rhines_scaling}
\end{align}
which is commonly referred to as the Rhines scaling \citep[e.g.][]{rhines_1975,cardin_and_olson_1994,aubert_etal_2001,cabanes_etal_2017,guervilly_etal_2019}.

\subsubsection{MAC balance}
\label{MAC_balance_theory}
The last option consists of assuming a triple balance between Lorentz, buoyancy and Coriolis terms in (\ref{vorticity_equation}), the so-called MAC balance (MAC for Magneto-Archimedean-Coriolis) \citep[e.g.][]{starchenko_and_jones_2002,davidson_2013,calkins_2018}. 
In addition to the aforementioned integral length scales $\mathcal{L}_{\parallel}$ and $\mathcal{L}_{\perp}$, the MAC balance theory put forward by  \cite{davidson_2013} (see also \cite{wicht_and_sanchez_2019} for a detailed derivation) introduces the magnetic dissipation length scale
\begin{align}
\mathcal{L}_{\mathrm{ohm}} = \sqrt{\frac{\int_{V} \mathbf{B}^{2} \, \mathrm{d} V}{\int_{V} \left( \nabla \times \mathbf{B} \right)^{2} \, \mathrm{d} V} },
\end{align}
where $V$ is the outer core volume. Ohmic dissipation is expected to carry most of the energy loss in the limit of magnetic energy much larger than kinetic energy ($E_{\mathrm{mag}} \gg E_{\mathrm{kin}}$) and small magnetic Prandtl number ($Pm = \nu / \lambda~\ll~1$). Both requirements are fulfilled in the Earth's core.

Assuming the ageostrophic flow dynamics described by (\ref{vorticity_equation}) to be controlled by a MAC balance, and using $\mathbf{j} \times \mathbf{B} \sim B^{2}/ \mu \mathcal{L}_{\perp}$, yields the following order of magnitude estimates
\begin{align}
\frac{B^{2}}{\rho \mu \mathcal{L}_{\perp}^{2}} \sim \frac{\alpha \Theta g_{\mathrm{o}}}{\mathcal{L}_{\perp}} \sim \frac{\Omega U}{\mathcal{L}_{\parallel}}.
\label{MAC_estimates}
\end{align} 
Combining the balance between the Lorentz and buoyancy terms with the estimate for the input power per unit mass 
\begin{align}
\mathcal{P} = \frac{\alpha g_{o}}{V r_{o}} \int_{V} r u_{r} T \mathrm{d} V \ \sim \ \alpha g_{o} U \Theta,
\label{power_per_mass}
\end{align}
yields the following relation for the magnetic energy density
\begin{align}
\frac{B^{2}}{\rho \mu} \ \sim \ \frac{\mathcal{P} \mathcal{L}_{\perp}}{U}.
\label{magnetic_energy_per_mass_1}
\end{align}
In the Earth's core, the energy input is expected to be almost entirely balanced by ohmic dissipation
\begin{align}
\mathcal{D}_{\mathrm{ohm}} = \frac{1}{\rho V} \int_{V} \frac{\lambda}{\mu} \left( \nabla \times \mathbf{B} \right)^{2} \mathrm{d}V.
\label{ohmic_dissipation}
\end{align}
However, in numerical dynamos this may not necessarily be the case due to viscosity still being sizeable. To account for remaining viscous effects, the factor $f_{\mathrm{ohm}}$, which quantifies the relative fraction of heat dissipated via Ohmic losses, is introduced. As a result, the total energy loss per unit mass can be expressed by
\begin{align}
\mathcal{D} = \frac{1}{f_{\mathrm{ohm}}} \mathcal{D}_{\mathrm{ohm}} \ \sim \ \frac{1}{f_{\mathrm{ohm}}} \frac{\lambda B^{2}}{\mu \rho \mathcal{L}_{\mathrm{ohm}}^{2}}.
\label{total_disspation_per_mass}
\end{align}
Hence, considering that for a saturated dynamo the energy input should be balanced by dissipation, $\mathcal{P} \sim \mathcal{D}$, yields
\begin{align}
\frac{B^{2}}{\rho \mu} \ \sim \ \frac{\mathcal{L}_{\mathrm{ohm}}^{2}}{\lambda} f_{\mathrm{ohm}} \mathcal{P}.
\label{magnetic_energy_per_mass_2}
\end{align}
By combining equations (\ref{magnetic_energy_per_mass_1}) and (\ref{magnetic_energy_per_mass_2}), we therefore obtain
\begin{align}
f_{\mathrm{ohm}} \frac{U}{\mathcal{L}_{\perp}} \ \sim \ \frac{\lambda}{\mathcal{L}_{\mathrm{ohm}}^{2}}.
\label{vorticity_equivalence}
\end{align}
In the above expression, $U / \mathcal{L}_{\perp}$ denotes the typical large-scale vorticity related to the convective columns, while $\lambda / \mathcal{L}_{\mathrm{ohm}}^{2}$ can be interpreted as a characteristic small-scale vorticity \citep[see][]{davidson_2013}. Thus, when $f_{\mathrm{ohm}} = 1$, Eq. (\ref{vorticity_equivalence}) indicates that large- and small-scale vorticities are proportional to each other. This proportionality is an inherent feature of quasi-2D turbulence such as rapidly-rotating convection \citep[e.g.][]{davidson_2004}. The non-dimensional form of Eq. (\ref{vorticity_equivalence}) provides a way to estimate the integral length scale~$\mathcal{L}_{\perp}$
\begin{align}
\frac{\mathcal{L}_{\perp}}{L} \ \sim \ f_{\mathrm{ohm}} \ \left(\frac{\mathcal{L}_{\mathrm{ohm}}}{L}\right)^{2} Rm,
\label{vorticity_equivalence_nondim}
\end{align}
where $Rm = U L / \lambda$ is the magnetic Reynolds number. 
Making the additional assumption that large- and small-scale vorticities are independent of the rotation rate implies that the magnetic energy density itself is also independent of $\Omega$. Based on a dimensional analysis, \cite{davidson_2013} argues that $B^{2}/\rho \mu$ can hence be expressed as a function of the power per mass generated by buoyancy forces and the flow length scale alone, which yields
\begin{align}
\frac{B^{2}}{\rho \mu} \ \sim \ \mathcal{L}_{\parallel}^{2/3} \left( f_{\mathrm{ohm}} \mathcal{P} \right)^{2/3}.
\label{Bsquare_power}
\end{align}
The balance between the buoyancy and Coriolis terms in (\ref{MAC_estimates}) together with Eq. (\ref{power_per_mass}) results in the following relation
\begin{align}
\frac{\mathcal{L}_{\perp}}{\mathcal{L}_{\parallel}} \ \sim \ \frac{\mathcal{P}}{\Omega U^{2}}.
\label{flow_length_scale}
\end{align}
Combining Eqs. (\ref{magnetic_energy_per_mass_1}), (\ref{Bsquare_power}) and (\ref{flow_length_scale}), and assuming $\mathcal{L}_{\parallel} \sim L$, leads to the following scaling for the flow length scale
\begin{align}
\frac{\mathcal{L}_{\perp}}{L} \ \sim \ f_{\mathrm{ohm}}^{1/2} Ro^{1/4}.
\label{Rossby_one_fourth}
\end{align}
Therefore, we finally end up with an expression that only depends on a system-scale dimensionless number.

\subsection{Leading-order magnetostrophy}
As an alternative to a prevalent geostrophic balance, it has been suggested that the flow dynamics in the Earth's core could be in a magnetostrophic regime \citep[e.g.][]{roberts_1978, hollerbach_1996}. This would imply that the leading-order force balance consists of Coriolis, pressure and Lorentz forces
\begin{align}
2 \rho \mathbf{\Omega} \times \mathbf{u} =  -\nabla P + \mathbf{j} \times \mathbf{B}.
\end{align}
In this scenario, the Lorentz force is expected to be strong enough to relax the rotational constraint imposed by the Coriolis force. It has been commonly assumed that the convective length scale can then reach the system scale
\begin{align}
\mathcal{L}_{\parallel} \sim \mathcal{L}_{\perp} \sim L.
\end{align}
So far, system-scale magnetostrophy has not been attained in numerical geodynamo simulations, although it has been approached in some models with strong magnetic control \citep[e.g.][]{dormy_2016,raynaud_etal_2020}. Numerical dynamos in the currently accessible parameter space, however, frequently exhibit a magnetostrophic balance at smaller length scales \citep[e.g.][]{aurnou_and_king_2017,aubert_2019}.

\label{elsasser_scaling}

The assessment of whether a dynamo is in a magnetostrophic balance is generally based on the evaluation of the Elsasser number $\Lambda$, which is a measure of the relative strengths of the Lorentz and Coriolis forces:
\begin{align}
\Lambda = \frac{|\mathbf{F}_{\mathrm{Lorentz}}|}{|\mathbf{F}_{\mathrm{Coriolis}}|} = \frac{|\mathbf{j} \times \mathbf{B}|}{|2 \rho \Omega \times \mathbf{u}|} \ \sim \  \frac{JB}{\rho \Omega U}.
\label{elsasser_number}
\end{align}
Using Ohm's law to scale the current density as $J \sim \sigma U B$, where $\sigma = 1 / \left( \mu \lambda \right)$, results in the traditional form of the Elsasser number
\begin{align}
\Lambda_{\mathrm{t}} = \frac{B^{2}}{\rho \mu \lambda \Omega}
\end{align}
This definition allows one to obtain estimates of the force ratio for planetary dynamos based on magnetic field observations, yielding $\Lambda_{\mathrm{t}} \sim \mathcal{O} \left(10\right)$ for the Earth. This has been used to argue for a dominant role of Lorentz forces, and therefore the geodynamo being in the magnetostrophic regime. \citet{soderlund_etal_2012} note that this might not be the correct interpretation of $\Lambda_{\mathrm{t}}$ since the estimate $J \sim \sigma U B$ is only expected to hold when $Rm \ll 1$. This is, however, not the case for global-scale dynamics in the Earth's core for which $Rm \sim  \mathcal{O} \left(10^{3}\right)$. Hence, $\Lambda_{\mathrm{t}}$ is only relevant on small length scales where $Rm \ll 1$ and is likely not an appropriate measure to assess global-scale magnetostrophy. \cite{soderlund_etal_2012,soderlund_etal_2015} suggested that a more meaningful estimate of the relative strengths of Lorentz and Coriolis forces in the Earth's core can be obtained by scaling the electric current density using Amp\`ere's law under the magneto-hydrodynamic approximation
\begin{align}
\mathbf{J} =  \frac{\nabla \times \mathbf{B}}{\mu} \ \sim \ \frac{B}{\mu \mathcal{L}_{B}},
\end{align}
where $\mathcal{L}_{B}$ is the length scale of magnetic field structures.
Inserting this into (\ref{elsasser_number}) results in the dynamic Elsasser number \citep[e.g.~][]{christensen_etal_1999,cardin_etal_2002,soderlund_etal_2012}
\begin{align}
\Lambda_{\mathrm{d}} = \frac{B^{2}}{\rho \mu \Omega U \mathcal{L}_{B}} \ \sim \ \frac{\Lambda_{\mathrm{t}}}{Rm} \frac{L}{\mathcal{L}_{B}}.
\label{dynamic_elsasser_number}
\end{align}
The dependence of $\Lambda_{\mathrm{d}}$ on $\mathcal{L}_{B}^{-1}$ implies that the relative strength of the Lorentz force increases with decreasing length scale. Building on these developments, \cite{aurnou_and_king_2017} attempted to estimate the flow scale at which the zeroth-order force balance changes from geostrophy to magnetostrophy, that is the flow scale at which $\Lambda_{\mathrm{d}} = 1$. To this end, they assumed quasi-steady induction ($\partial \mathbf{B} / \partial t = 0$) which results in the following balance:
\begin{align}
\frac{BU}{\mathcal{L}_{U}} \ \sim \ \frac{\lambda B}{\mathcal{L}_{\mathrm{B}}^{2}},
\end{align}
where $\mathcal{L}_{U}$ is the length scale of the flow.
This yields the following relation for the magnetic length scale:
\begin{align}
\frac{\mathcal{L}_{B}}{L} \ \sim \ \left( \frac{\mathcal{L}_{U}}{Rm L} \right)^{1/2}.
\label{nondim_magnetic_scale}
\end{align}
Replacing $\mathcal{L}_{B}/L$ in (\ref{dynamic_elsasser_number}) with relation (\ref{nondim_magnetic_scale}) results in
\begin{align}
\Lambda_{\mathrm{d}} \ \sim \ \left( \frac{\Lambda_{\mathrm{t}}^{2}}{Rm} \frac{L}{\mathcal{L}_{U}} \right)^{1/2}.
\end{align}
Based on this expression \cite{aurnou_and_king_2017} interpret $\Lambda_{\mathrm{t}}^{2}/Rm$ as a dimensionless length scale
\begin{align}
\frac{\mathcal{L}_{X}}{L} = \frac{\Lambda_{\mathrm{t}}^{2}}{Rm}.
\label{magnetostrophic_crossover}
\end{align}
The dynamic Elsasser number can therefore be rewritten as
\begin{align}
\Lambda_{\mathrm{d}} \ \sim \ \left( \frac{\mathcal{L}_{X}}{\mathcal{L}_{U}} \right)^{1/2},
\end{align}
indicating that the flow is in geostrophic balance on length scales larger than $\mathcal{L}_{X}$ due to the Lorentz force being subdominant to the Coriolis force ($\Lambda_{\mathrm{d}} < 1$). On scales smaller than $\mathcal{L}_{X}$, the Lorentz force is expected to be dominant ($\Lambda_{\mathrm{d}} > 1$), suggesting that the flow is in magnetostrophic balance. Hence, \cite{aurnou_and_king_2017} refer to $\mathcal{L}_{X}$ as the magnetostrophic cross-over length scale.

\section{Methods}
\label{methods}
\subsection{Numerical models}
For the present study, we extend the set of numerical dynamo models analysed by \cite{schwaiger_etal_2019} with non-magnetic rotating convection models.
In both sets of simulations, we consider a spherical shell rotating about the axis $\mathbf{e}_{z}$ with constant angular frequency $\Omega$, with a ratio between the inner and outer radii of  $r_{\mathrm{i}} / r_{\mathrm{o}} = 0.35$. The shell is filled with an incompressible fluid of density $\rho$ and kinematic viscosity $\nu$, which is electrically conducting in the dynamo models and electrically insulating in the rotating convection models. 
An imposed temperature difference $\Delta T = T_{\mathrm{o}}-T_{\mathrm{i}}$ between the two bounding spheres drives convection of the fluid. In addition to being held at constant temperatures, both boundaries are mechanically rigid and electrically insulating.

We solve the dimensionless magneto-hydrodynamic equations under the Boussinesq approximation for the velocity field $\mathbf{u}$, magnetic field $\mathbf{B}$ and temperature $T$. The shell depth $L=r_{\mathrm{o}}-r_{\mathrm{i}}$ serves as the reference length scale and the viscous diffusion time $L^{2}/\nu$ is the time unit. The temperature is scaled by $\Delta T$, and the magnetic field by $\sqrt{\rho \mu \lambda \Omega}$, where $\mu$ is the magnetic permeability and $\lambda$ the magnetic diffusivity. The dimensionless gravity profile is assumed to be linear and follows $g(r)=r/r_{\mathrm{o}}$. Hence, we end up with the following system of equations:
\begin{align}
\begin{split}
   \frac{\partial \mathbf{u}}{\partial t}+\mathbf{u}\cdot \nabla \mathbf{u} + \frac{2}{E} \mathbf{e}_{z}\times \mathbf{u} = &- \nabla P + \frac{Ra}{Pr}\frac{\mathbf{r}}{r_{o}}T + \nabla^{2}\mathbf{u} \ + \\ & + \frac{1}{E Pm}\left(\nabla \times \mathbf{B} \right)\times \mathbf{B}, 
\end{split} \label{nondim_Navier_Stokes_equation}\\[5pt]
   \frac{\partial T}{\partial t} + \mathbf{u} \cdot \nabla T =& \ \frac{1}{Pr}\nabla^{2}T,\\[4pt]
   \frac{\partial \mathbf{B}}{\partial t} =& \ \nabla \times \left( \mathbf{u}\times \mathbf{B} \right) + \frac{1}{Pm} \nabla^{2}\mathbf{B} , \label{induction_equation}\\[4pt]
   \nabla \cdot \mathbf{u}  =& \ 0,\\[5pt]
   \nabla \cdot \mathbf{B} =& \ 0 \label{solenoidal_B_equation},
\end{align}
where $P$ corresponds to the dimensionless pressure. The control parameters governing this set of equations are the Ekman number
\begin{align}
E = \frac{\nu}{\Omega L^{2}},
\end{align}
the hydrodynamic Prandtl number
\begin{align}
Pr = \frac{\nu}{\kappa},
\end{align}
the magnetic Prandtl number
\begin{align}
Pm = \frac{\nu}{\lambda}
\end{align}
and the Rayleigh number
\begin{align}
Ra = \frac{\alpha g_{o} L^{3} \Delta T}{\nu \kappa},
\end{align}
where $\kappa$ is the thermal diffusivity, $\alpha$ the thermal expansion coefficient and $g_{\mathrm{o}}$ the gravity at the outer boundary.

All simulations considered in this study were computed using the open-source code MagIC \citep[][freely available at \url{https://github.com/magic-sph/magic}]{wicht_2002, gastine_etal_2016}. To numerically solve Eqs. (\ref{nondim_Navier_Stokes_equation}-\ref{solenoidal_B_equation}) in the spherical coordinate system $\left( r, \theta, \varphi \right)$,
the solenoidal vector fields $\mathbf{u}$ and $\mathbf{B}$ are decomposed into poloidal and toroidal potentials
\begin{align}
\mathbf{u} = \nabla \times \nabla \times \left( W \mathbf{e}_{r} \right) + \nabla \times \left( Z \mathbf{e}_{r} \right), \\
\mathbf{B} = \nabla \times \nabla \times \left( G \mathbf{e}_{r} \right) + \nabla \times \left( H \mathbf{e}_{r} \right),
\end{align}
where $\mathbf{e}_{r}$ is the radial unit vector. The spatial discretisation of the unknown scalar fields $W$, $Z$, $G$, $H$, $T$ and $P$ involves a spherical harmonic expansion up to degree and order $\ell_{\mathrm{max}}$ in the angular directions, and a Chebyshev decomposition with $N_{r}$ collocation points in the radial direction. MagIC employs the open-source libary SHTns \citep[][freely available at \url{https://bitbucket.org/nschaeff/shtns}]{schaeffer_2013} for efficient computation of the spherical harmonic transforms. The equations are integrated in time using a semi-implicit adaptive time stepping algorithm. The Coriolis force and the non-linear terms are treated explicitly using a second-order Adams-Bashforth scheme, while the remaining terms are advanced implicitly with a Crank-Nicolson scheme.

To investigate the link between force balances and observable flow length scales, we analyse 95 dynamo models and 24 non-magnetic rotating convection models. 
The set of dynamo models is nearly identical to the one studied by \citet[][Table A1]{schwaiger_etal_2019}, with the only difference being that the run time of some of the simulations has been increased to improve the statistics of the output parameters. The control parameters and relevant results of the non-magnetic cases can be found in Table \ref{appendix_hydro_table}.

\subsection{Energetically relevant length scales}
In numerical dynamo and rotating convection models, characteristic length scales of the convective flow are typically obtained from the kinetic energy spectra.
In our study, we consider the time-averaged spectrum of the poloidal kinetic energy, which is defined by \citep[see][p. 159]{glatzmaier_2013}
\begin{align}
\mathcal{E}_{\mathrm{pol}} = \sum_{\ell=0}^{\ell_{\mathrm{max}}} \mathcal{E}_{\mathrm{pol}, \ell},
\end{align}
where
\begin{align}
\begin{split}
\mathcal{E}_{\mathrm{pol}, \ell} = \int_{r_{\mathrm{i}}}^{r_{\mathrm{o}}} \sideset{}{'}\sum_{m=0}^{\ell}  \ell \left( \ell + 1 \right) \left[ \vphantom{\left| \frac{\partial W_{\ell}^{m}}{\partial r} \right|^{2}} \frac{\ell \left( \ell + 1 \right)}{r^{2}} \left| W_{\ell}^{m} \right|^{2} \right. + \\ \left. + \left| \frac{\partial W_{\ell}^{m}}{\partial r} \right|^{2} \right] \, \mathrm{d} r.
\end{split}
\end{align}
In the above expression $W_{\ell}^{m}$ is the poloidal flow potential of degree $\ell$ and order $m$. The prime on the summation indicates that the $m=0$ contribution entering the sum is multiplied by one half. We choose the peak of the spectrum of $\mathcal{E}_{\mathrm{pol}}$  to characterise the convective pattern of the flow in our simulations, i.e.
\begin{align}
\ell_{\mathrm{pol}} = \argmax_{\ell} \left( \mathcal{E}_{\mathrm{pol}, \ell}  \right).
\end{align}
The degree $\ell_{\mathrm{pol}}$ can be associated to a length scale $\mathcal{L}_{\mathrm{pol}}$ through the definition of the characteristic half-wavelength \citep[see e.g.][p.~101]{backus_etal_1996}
\begin{align}
\mathcal{L}_{\mathrm{pol}} = \frac{\pi r_{m}}{\sqrt{\ell_{\mathrm{pol}} \left( \ell_{\mathrm{pol}} + 1 \right)}} \approx \frac{\pi r_{m}}{\ell_{\mathrm{pol}}},
\label{L_pol}
\end{align}
where the mid-shell radius $r_{m} = (r_{\mathrm{i}}+r_{\mathrm{o}})/2$ is approximated by $1$ when $r_{\mathrm{i}} / r_{\mathrm{o}} = 0.35$. As we shall see in Fig.~\ref{Figure_MAC_scale}a, $\mathcal{L}_{\mathrm{pol}}$ enables the recovery of results previously obtained with the more commonly used energy-weighted length scale \citep{christensen_and_aubert_2006}, while being more representative of the dominant scale of convection and arguably less sensitive to second-order force balances that would control the tail of the spectrum \citep[e.g.][]{aubert_etal_2017,dormy_etal_2018}.

\begin{figure*}
\centering
\includegraphics[width=1\textwidth]{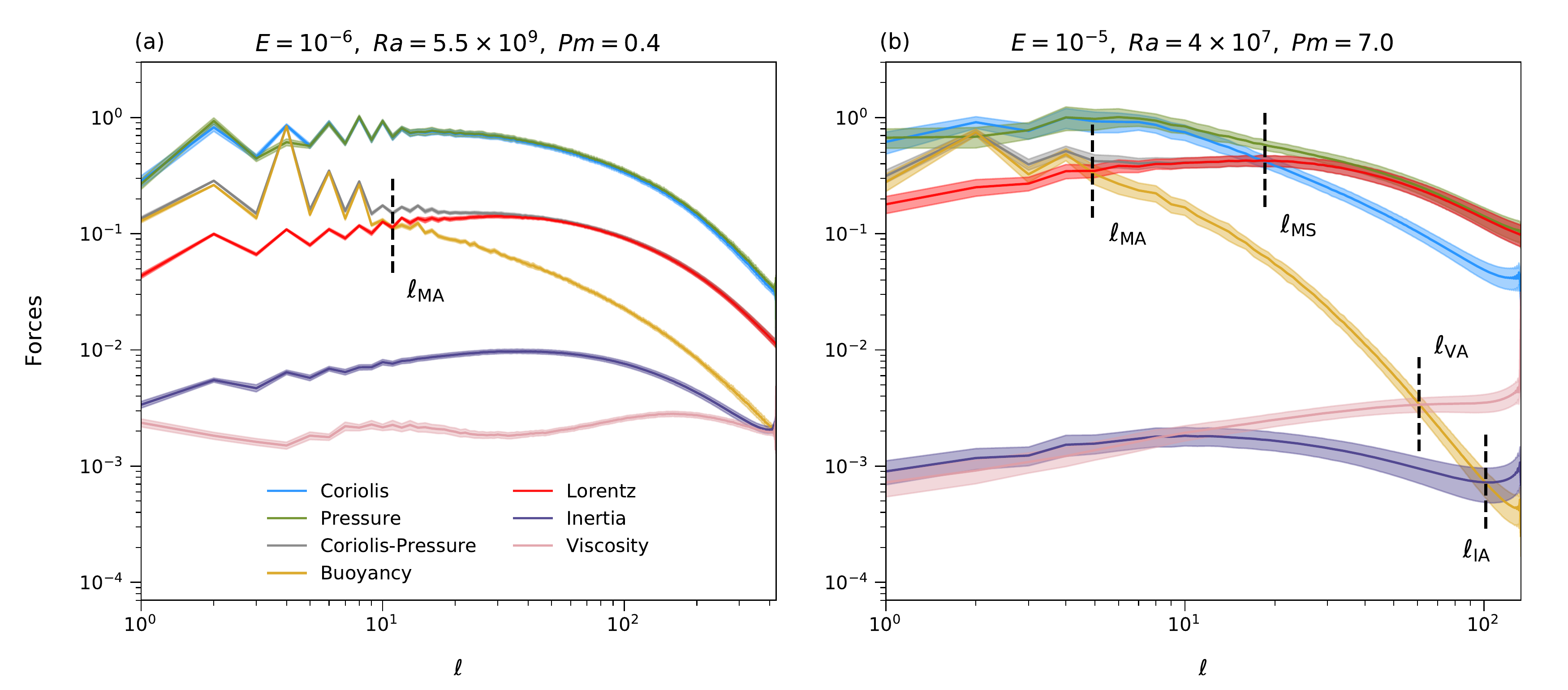}
\caption[]{Examples of time-averaged force balance spectra for two strong-field QG-MAC dynamos. The amplitudes of the spherical harmonic contributions of the forces are normalised relative to the peak of the Coriolis force. The shaded regions correspond to one standard deviation in time. (a) Force balance of a  dynamo ($M = E_{\mathrm{mag}}/E_{\mathrm{kin}} \approx 10$) governed by a prevailing geostrophic balance at all length scales, followed by a first-order MAC balance. (b) Force balance of a dynamo ($M \approx 200$) with a subdivided zeroth-order force balance that is controlled by a geostrophic balance at large scales (small $\ell$) that transitions into a magnetostrophic balance towards smaller scales (large $\ell$).}
\label{Figure_mag_force_balances}
\end{figure*}

\subsection{Dynamically relevant length scales}
To examine whether the energetically relevant flow length scales can be related to the governing force balances, we define dynamically relevant length scales, i.e. length scales that are representative of the underlying force equilibria. To this end, we rely on the spectral representation of the force balance introduced by \citet{aubert_etal_2017}. This requires an expression of each force vector~$\boldsymbol{f}$ in terms of scalar potentials, i.e.
\begin{align}
\boldsymbol{f} = \mathcal{R} \mathbf{e}_{r} + r \nabla \mathcal{S} + \mathbf{r} \times \nabla \mathcal{T}, 
\end{align}
where $\mathbf{r} = r \mathbf{e}_{r}$ is the radius vector. $\mathcal{R}$, $\mathcal{S}$ and $\mathcal{T}$ represent the radial, spheroidal and toroidal scalar fields, respectively. The latter three quantities can then be expanded in spherical harmonics, such that
\begin{align}
\boldsymbol{f} = \sum_{\ell=0}^{\ell_{\mathrm{max}}} \sum_{m=-\ell}^{\ell} \mathcal{R}_{\ell}^{m} Y_{\ell}^{m} \mathbf{e}_{r} + \mathcal{S}_{\ell}^{m} r \nabla Y_{\ell}^{m} + \mathcal{T}_{\ell}^{m} \mathbf{r} \times \nabla Y_{\ell}^{m}, 
\end{align}
where $Y_{\ell}^{m}$ are the spherical harmonic functions of degree $\ell$ and order $m$. The energy of the force vector (excluding viscous boundary layers) can then be obtained by computing
\begin{align}
\begin{split}
F^{2} \, & = \int_{V} \boldsymbol{f}^{2} \, \mathrm{d} V\\
& = 2 \int_{r_{\mathrm{i}} + b}^{r_{\mathrm{o}}-b} \sum_{\ell=0}^{\ell_{\mathrm{max}}} \sideset{}{'}\sum_{m=0}^{\ell} \left| \mathcal{R}_{\ell}^{m} \right|^{2} + \ell \left( \ell+1 \right) \left( \left| \mathcal{S}_{\ell}^{m} \right|^{2} + \left| \mathcal{T}_{\ell}^{m} \right|^{2} \right) r^{2} \, \mathrm{d} r, 
\end{split}
\end{align}
where $b$ is the thickness of the viscous boundary layers. The prime on the summation again indicates that the first term of the sum is multiplied by one half. The above expression can be rewritten as
\begin{align}
F^{2} = \sum_{\ell=0}^{\ell_{\mathrm{max}}} \mathcal{F}_{\ell}^{2},
\label{Force_sum}
\end{align}
where
\begin{align}
\mathcal{F}_{\ell}^{2} = 2 \int_{r_{\mathrm{i}} + b}^{r_{\mathrm{o}}-b} \sideset{}{'}\sum_{m=0}^{\ell} \left| \mathcal{R}_{\ell}^{m} \right|^{2} + \ell \left( \ell+1 \right) \left( \left| \mathcal{S}_{\ell}^{m} \right|^{2} + \left| \mathcal{T}_{\ell}^{m} \right|^{2} \right) r^{2} \, \mathrm{d} r.
\label{Force_spectra}
\end{align}
The above formalism can be used not only to compute spectra of individual forces, but also to measure the degree of cancellation between forces, such as between the pressure gradient and the Coriolis force, which yields the ageostrophic Coriolis force.

\begin{figure*}
\centering
\includegraphics[width=\textwidth]{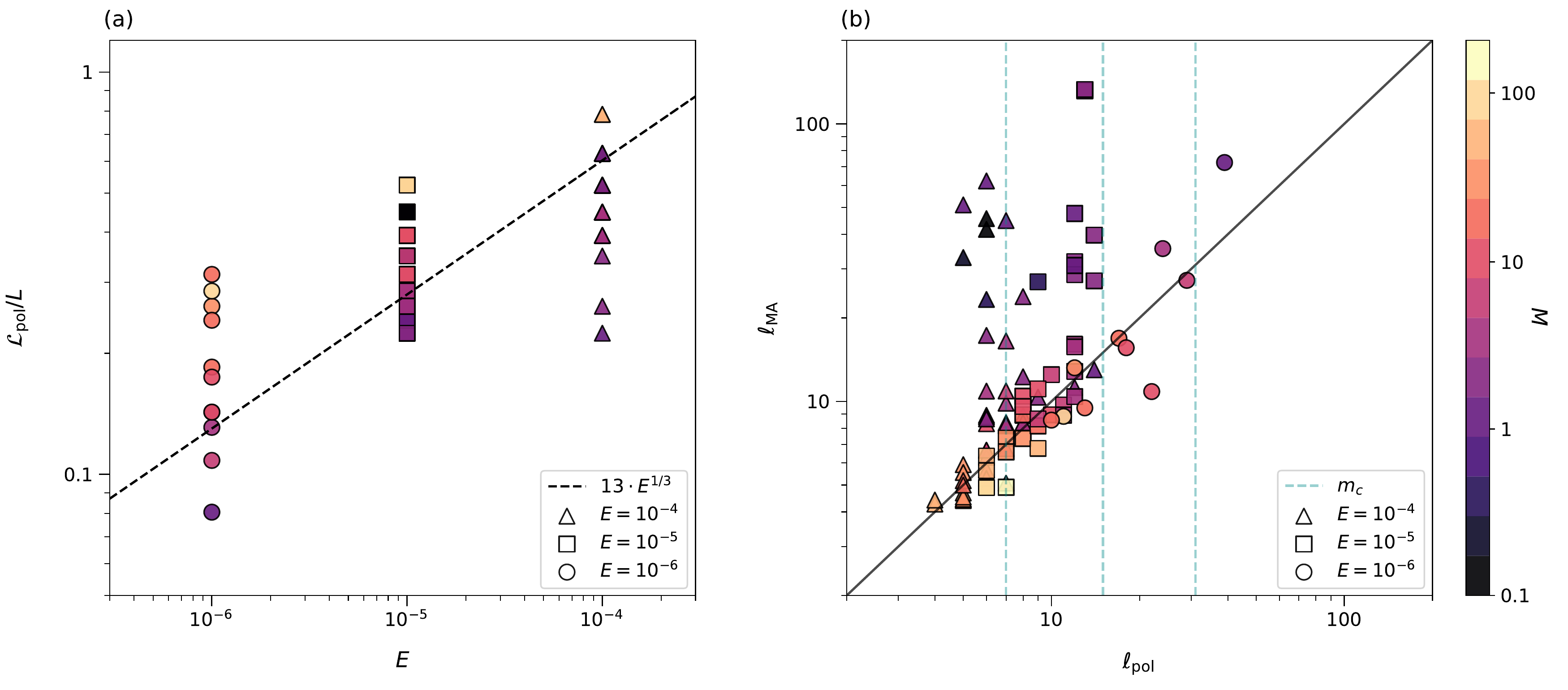}
\caption[]{(a) $\mathcal{L}_{\mathrm{pol}} / L$ (Eq. \ref{L_pol}) as a function of the Ekman number $E$. The black dashed line corresponds to the viscous scaling $\mathcal{L}_{\perp} / L \sim E^{1/3}$. (b) Comparison of the spherical harmonic degree at which buoyancy and Lorentz forces are of equal magnitude in the force balance spectra, $\ell_{\mathrm{MA}}$, (see Fig. \ref{Figure_mag_force_balances} for examples) to the peak of the poloidal kinetic energy spectrum $\ell_{\mathrm{pol}}$. The spherical harmonic degrees can be converted to length scales using Eq.(\ref{length_scale_conversion}). The green vertical dashed lines correspond to the respective azimuthal wavenumber of convection onset $m_{c}$ for $E \in \{ 10^{-4}, 10^{-5}, 10^{-6} \}$ from left to right. The symbols in both panels are coloured with $M$.}
\label{Figure_MAC_scale}
\end{figure*}

Figure \ref{Figure_mag_force_balances} illustrates the force spectra $\mathcal{F}_{\ell}$ (time-averaged and normalised by the peak of the spectrum of the Coriolis force) of two dynamos controlled by a QG-MAC balance. This type of force equilibrium has been shown to be the governing force balance of most dipole-dominated dynamos in a systematic parameter space survey by \cite{schwaiger_etal_2019}. At zeroth order these dynamos are controlled by a balance between pressure and Coriolis forces, the so-called geostrophic balance. 
However, depending on the strength of the magnetic field, the geostrophic balance is either present at all length scales (see Fig. \ref{Figure_mag_force_balances}a) or restricted to large scales, that is small spherical harmonic degrees (see Fig. \ref{Figure_mag_force_balances}b). In the latter case, the zeroth-order balance morphs into a balance between pressure and Lorentz forces towards smaller scales, forming a magnetostrophic balance.  
The Lorentz force contributes to this balance predominantly in the form of magnetic pressure. At the following order, the cancellation between pressure and Coriolis forces, the ageostrophic Coriolis force, is balanced by buoyancy at large scales (small $\ell$) and the Lorentz force at small scales (large $\ell$). This first-order balance is typically referred to as MAC balance. Finally, inertia and viscous forces contribute at second order. To indicate the ``strong-fieldness'' of the dynamo, i.e. the dynamical influence of the Lorentz force relative to the second-order forces, we will use the ratio between the magnetic and kinetic energies $M = E_{\mathrm{mag}} / E_{\mathrm{kin}}$ \citep[e.g.][]{schwaiger_etal_2019}. In the following, we will refer to cases with $M \geq 10$ as strong-field dynamos. 

In dynamo models, there are up to four possible cross-overs of forces (see Fig. \ref{Figure_mag_force_balances}), which can be linked to different types of force balances: MAC, CIA, VAC and MS (magnetostrophic). The types of crossings which can be observed in a given dynamo model directly depends on the relative strength of the individual forces. The spherical harmonic degrees $\ell_{\mathrm{MA}}$, $\ell_{\mathrm{IA}}$ and $\ell_{\mathrm{VA}}$, corresponding to first- or higher-order MAC, CIA and VAC force balances, represent the scales at which the Lorentz force, inertia and viscous forces are of the same amplitude as buoyancy, respectively. Thus, we define
\begin{align}
\ell_{\mathrm{MA}} &= \argmin_{\ell} \left( \vert \mathcal{F}_{\mathrm{Lorentz},\ell} - \mathcal{F}_{\mathrm{buoyancy},\ell} \vert \right), \\
\ell_{\mathrm{IA}} &= \argmin_{\ell} \left( \vert \mathcal{F}_{\mathrm{inertia},\ell} - \mathcal{F}_{\mathrm{buoyancy},\ell} \vert \right), \\
\ell_{\mathrm{VA}} &= \argmin_{\ell} \left( \vert \mathcal{F}_{\mathrm{viscous},\ell} - \mathcal{F}_{\mathrm{buoyancy}, \ell} \vert \right).
\end{align}
The magnetostrophic crossing $\ell_{\mathrm{MS}}$ occuring at zeroth order, represents the scale at which Lorentz and Coriolis forces are in balance. It is therefore given by
\begin{align}
\ell_{\mathrm{MS}} = \argmin_{\ell} \left( \vert \mathcal{F}_{\mathrm{Coriolis},\ell} - \mathcal{F}_{\mathrm{Lorentz},\ell} \vert \right).
\end{align}
The length scales associated to these crossings (the so-called cross-over length scales) are again given by the characteristic half-wavelength associated to each degree (see Eq. \ref{L_pol})
\begin{align}
\mathcal{L} = \frac{\pi r_{m}}{\ell}.
\label{length_scale_conversion}
\end{align}

\section{Results}
\label{results}
\subsection{Convective length scales in dynamo simulations}
In planetary dynamos, the flow length scale is expected to be controlled by the Lorentz force. Therefore, we will examine in the following sections whether it is possible to relate the QG-MAC length scale $\mathcal{L}_{\mathrm{MA}} / L =\pi/\ell_{\mathrm{MA}}$ and the magnetostrophic cross-over length scale $\mathcal{L}_{\mathrm{MS}} / L =\pi/\ell_{\mathrm{MS}}$ to observable (energetically relevant) flow length scales. In addition, we will also examine their scaling behaviour to assess the validity of the aforementioned theoretical scalings in our numerical models.

\subsubsection{QG-MAC length scale}

\begin{figure*}
\centering
\includegraphics[width=\textwidth]{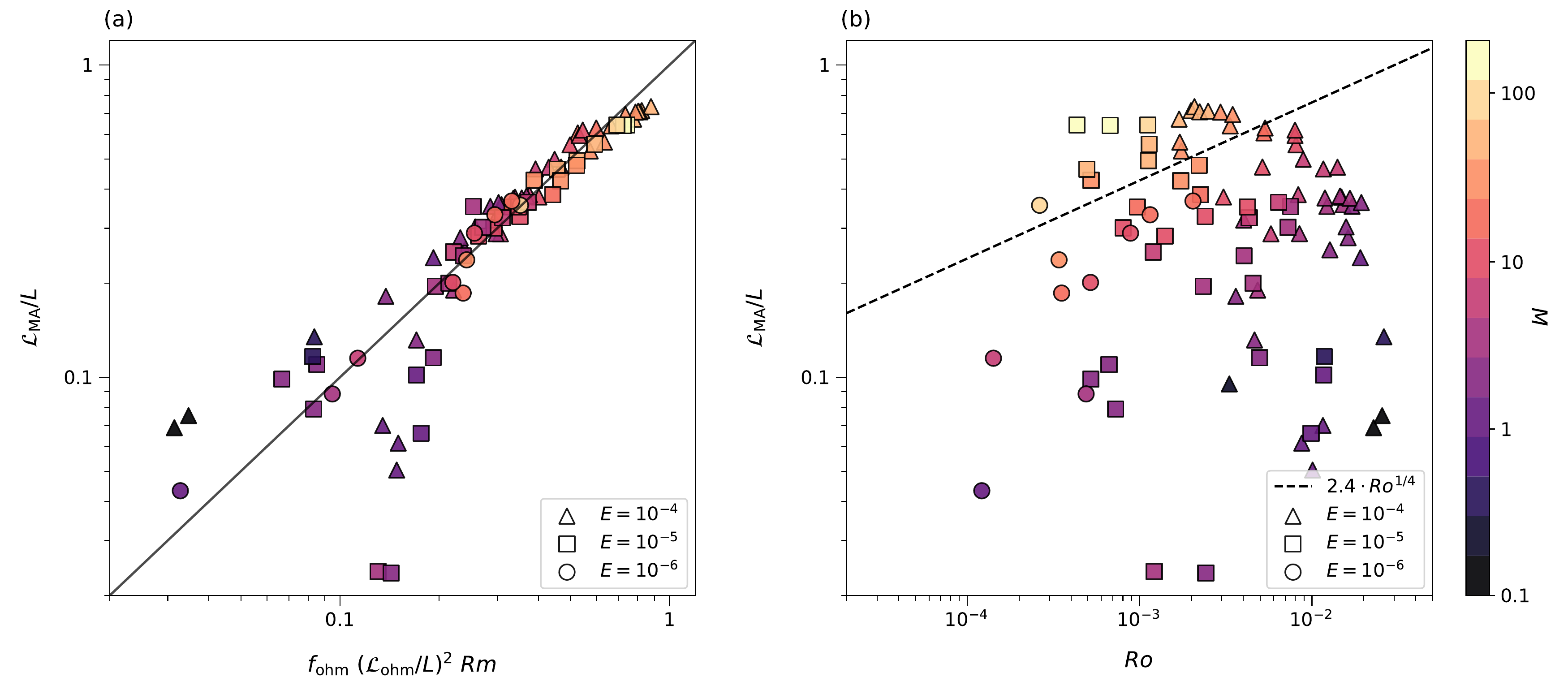}
\caption[]{(a) Comparison of the QG-MAC length scale $\mathcal{L}_{\mathrm{MA}}/L = \pi/\ell_{\mathrm{MA}}$ to the theoretical scaling given by Eq. (\ref{vorticity_equivalence_nondim}). (b) $\mathcal{L}_{\mathrm{MA}}/L$ as a function of $Ro$. The black dashed line corresponds to the theoretical scaling $\mathcal{L}_{\perp}/L \sim Ro^{1/4}$. The prefactor $2.4$ was obtained by  considering only dynamos with $M = E_{\mathrm{mag}}/E_{\mathrm{kin}} \geq 10$. The symbols in both panels are coloured with $M$.}
\label{Figure_MAC_scale_2}
\end{figure*}

One of the most frequently invoked results in favour of a QG-VAC balance in numerical dynamos is the apparent compatibility of a spectrally weighted kinetic energy length scale \citep{christensen_and_aubert_2006} with the viscous $E^{1/3}$-scaling \citep[e.g.][]{king_and_buffett_2013,oruba_and_dormy_2014}. Figure~\ref{Figure_MAC_scale}a shows the spectral peak length scale $\mathcal{L}_{\mathrm{pol}}$ as a function of the Ekman number. We observe that $\mathcal{L}_{\mathrm{pol}}$ also appears to follow a similar trend. This is however misleading, as the scatter of $\mathcal{L}_{\mathrm{pol}}$ at each given Ekman number is almost as large as the change in length scale predicted by the viscous scaling. This indicates that in this case simply considering the Ekman number as a diagnostic, without a more detailed analysis, has little to no predictive power for the underlying force balance.

The comparison of the dynamically relevant cross-over $\ell_{\mathrm{MA}}$ and the peak of the poloidal kinetic energy $\ell_{\mathrm{pol}}$ shown in Fig.~\ref{Figure_MAC_scale}b offers a more meaningful approach to link the dominant convective length scale to the governing physics. 
All models where $\ell_{\mathrm{MA}}$ is well defined are included independent of the type of the first-order force balance. To include information about the force balance of the dynamos, the symbols of the models are coloured with the ratio between the magnetic and kinetic energies $M$. We observe that $\ell_{\mathrm{MA}}$ is in good agreement with $\ell_{\mathrm{pol}}$ for dynamos with sufficiently large $M$ ($M \geq 10$),  which suggests that the convective scale is indeed controlled by a first-order MAC balance in these models. For dynamos with $1 < M < 10$ this correlation becomes less obvious due to the QG-MAC balance being increasingly perturbed by viscous and/or inertial effects. Finally, we observe no correlation between $\ell_{\mathrm{MA}}$ and $\ell_{\mathrm{pol}}$ for dynamos with $M \leq 1$, as expected since these models are no longer controlled by a QG-MAC balance.
The influence of second-order forces on the convective scale in dynamos with $M < 10$ is further highlighted by the fact that for each of the considered Ekman numbers, $\ell_{\mathrm{pol}}$ seems to cluster around the respective critical azimuthal wavenumber of convection onset (green vertical dashed lines in Figure \ref{Figure_MAC_scale}b), which follow the viscous scaling (Eq. \ref{viscous_scaling}). The onset values have been computed using the open-source generalised eigenvalue solver \textit{Singe} \citep[][freely available at \url{https://bitbucket.org/nschaeff/singe}]{vidal_and_schaeffer_2015}.

Given that $\mathcal{L}_{\mathrm{MA}}$ seems to correspond to an observable flow length scale in QG-MAC dynamos, we will now examine whether it is consistent with the theoretical predictions based on the assumption of a first-order MAC balance as suggested by \cite{davidson_2013} (see Section \ref{MAC_balance_theory}). 
The comparison between $\mathcal{L}_{\mathrm{MA}}$ and the prediction of the integral length scale $\mathcal{L}_{\perp}$ obtained from the equivalence between large- and small-scale vorticities (Eq. \ref{vorticity_equivalence_nondim}) is shown in Fig.~\ref{Figure_MAC_scale_2}a. 
The symbols corresponding to the individual models are again coloured with $M$. We observe that for numerical dynamos with strong magnetic control ($M \geq 10$), i.e. QG-MAC dynamos, the predicted and measured length scales are overall in good agreement. This suggests that the vorticity equivalence is reasonably well satisfied for these cases. The cross-over length scale $\mathcal{L}_{\mathrm{MA}}$ thus appears to correspond to the integral length scale $\mathcal{L}_{\perp}$ as defined by \cite{davidson_2013}.
This is again much less the case for dynamos with smaller $M$ since those dynamos either feature a significant contribution of inertial/viscous effects or are controlled by a different first-order force balance.

The second scaling of $\mathcal{L}_{\perp}$ (Eq. \ref{Rossby_one_fourth}) by \citet{davidson_2013} which further assumes the rotational independence of the vorticities is shown in Fig.~\ref{Figure_MAC_scale_2}b, in which $\mathcal{L}_{\mathrm{MA}}/L$ is plotted versus $Ro$. However, no correlation is found between $\mathcal{L}_{\mathrm{MA}}/L$ and the theoretical $\mathcal{L}_{\perp}/L \sim Ro^{1/4}$ scaling, even in the limit of $M \geq 10$.
Including the factor $f_{\mathrm{ohm}}^{1/2}$ in Eq. (\ref{Rossby_one_fourth}) does  not improve the correlation between the two quantities, which is why it is omitted in Fig. \ref{Figure_MAC_scale_2}b. There are several possible explanations for this. One of them being that our numerical models are not operating in a regime of $Pm \ll 1$, the physical limit in which the scaling~(\ref{Rossby_one_fourth}) was devised. Therefore, viscous dissipation is still significant (although some of the very high $Pm$ runs do have large $f_{\mathrm{ohm}}$). Additionally, the assumption of $B^{2} / \rho \mu $ being independent of the rotation rate, which is required for this scaling, is potentially not well satisfied in the models \citep{schrinner_2013}. In Fig.~\ref{Figure_MAC_scale_2}b, we also observe that the data points show a dependency on $M$, with larger length scales typically corresponding to larger $M$. 
This hints at a feedback between the magnetic field and the velocity field, which suggests that scaling the Lorentz force as $B^{2}/\mu \mathcal{L}_{\perp}$ is likely  too crude an approximation. While there are a number of reasons that could lead to the scaling law not being applicable to numerical dynamos, it should also be noted that in our models in which $M \geq 10$, $Ro$ only covers slightly more than one order of magnitude, which might not be enough for a fair assessment of its validity.

\subsubsection{Magnetostrophic cross-over length scale}
\begin{figure}
\centering
\includegraphics[width=0.45\textwidth]{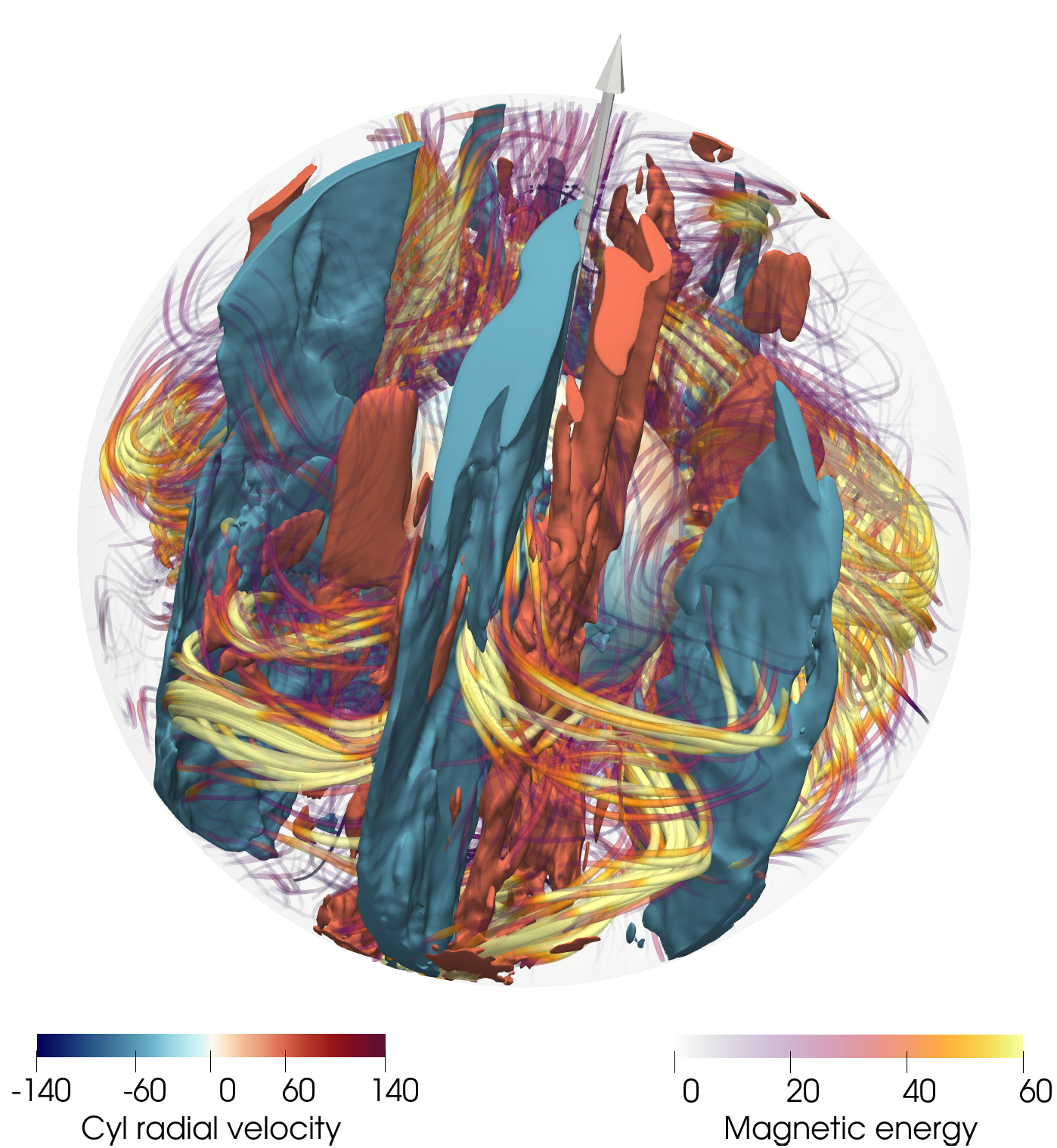}
\caption[]{3D-rendering of the columnar convection pattern and the magnetic field lines in a strong-field dynamo at $E=10^{-5}$, $Ra=4\times 10^{7}$ and $Pm=7$. The columns aligned with the rotation axis (represented by the arrow) correspond to isosurfaces of the cylindrical radial velocity $u_{s} \in \{-60,60\}$. The magnetic field lines are coloured according to the magnetic energy. Additionally, their thickness increases with the magnetic energy. The force balance corresponding to this case is shown in Fig.~\ref{Figure_mag_force_balances}b.}
\label{Figure_Flow_visualisation}
\end{figure}

In many strong-field dynamos the zeroth-order geostrophic balance is only present at large scales, while it turns into a balance between pressure and Lorentz forces towards smaller scales (see Fig. \ref{Figure_mag_force_balances}b). This transition from geostrophy to magnetostrophy occurs at the magnetostrophic cross-over length scale at which Lorentz and Coriolis forces are of equal magnitude \citep{aurnou_and_king_2017}.
The effect of this dichotomy of the zeroth-order force balance is visualised in Fig. \ref{Figure_Flow_visualisation} for one parameter configuration considered by \cite{dormy_etal_2018}.
We observe that the flow is organised into large-scale columns (spanning the entire shell) that are aligned with the rotation axis, indicating a dominant geostrophic balance on large scales. On smaller scales the columns are broken up especially in regions with an intense magnetic field.

\begin{figure*}
\begin{minipage}{0.33\textwidth}
\centering
\includegraphics[width=0.95\textwidth]{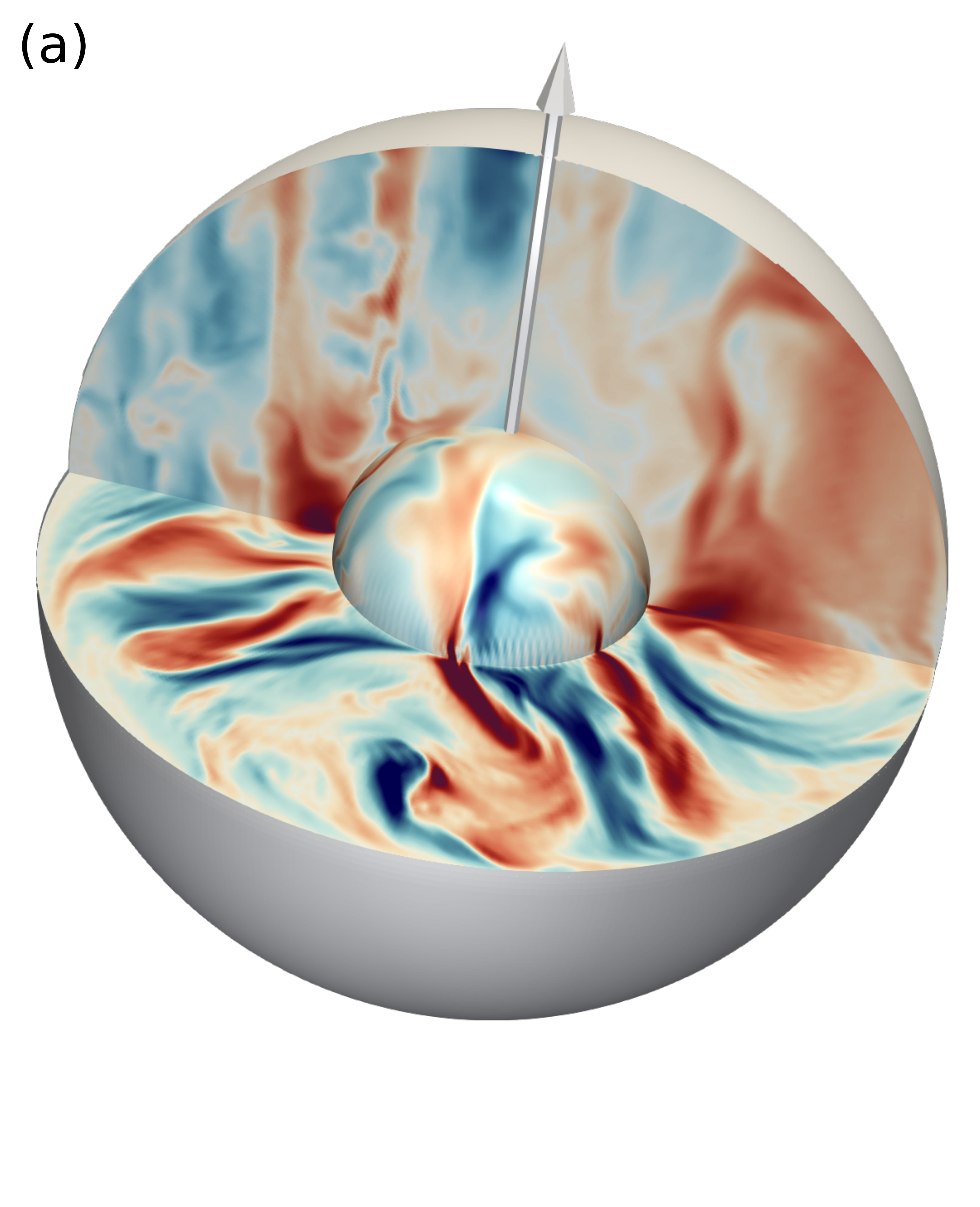}
\end{minipage}
\begin{minipage}{0.33\textwidth}
\centering
\includegraphics[width=0.95\textwidth]{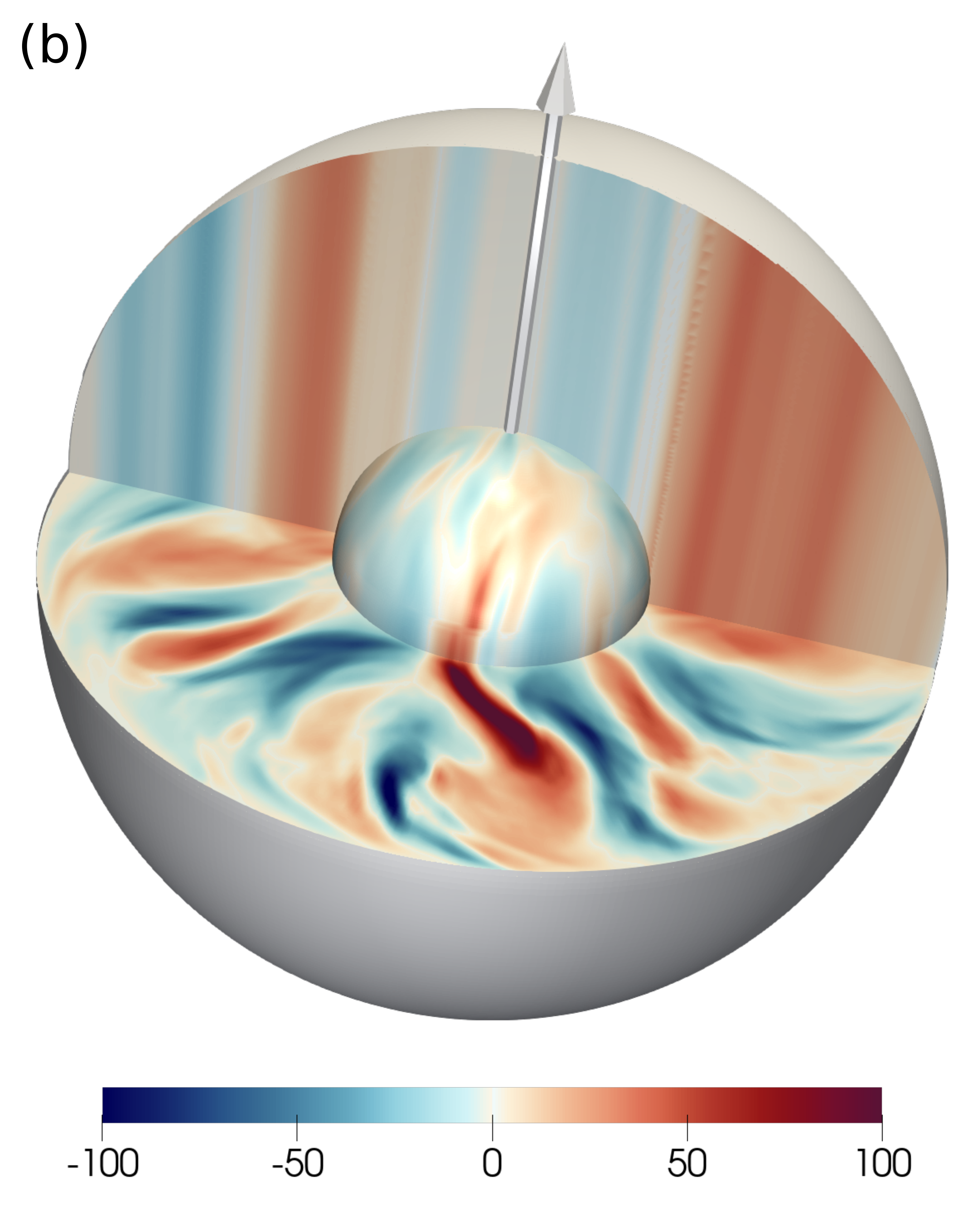}
\end{minipage}
\begin{minipage}{0.33\textwidth}
\centering
\includegraphics[width=0.95\textwidth]{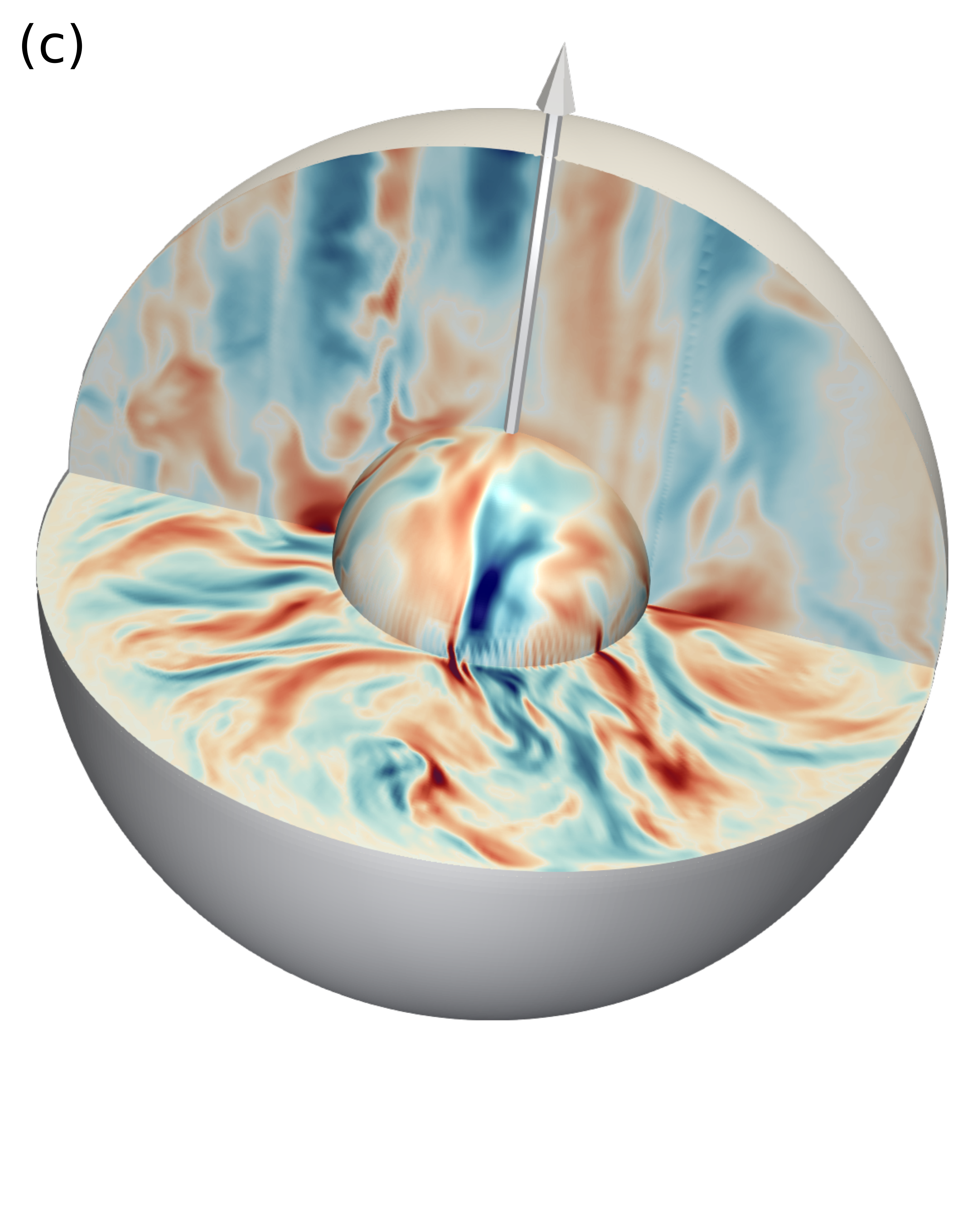}
\end{minipage}
\caption[]{3-D renderings of (a) the total cylindrical radial velocity $u_{s}$, (b) the geostrophic component $u_{s}^{g}$ of $u_{s}$, and (c) the ageostrophic component $u_{s}^{a}$ of $u_{s}$ for the numerical dynamo at $E=10^{-5}$, $Ra=4\times 10^{7}$ and $Pm=7$. In each panel, the arrow marks the rotation axis. The force balance corresponding to this case is shown in Fig.~\ref{Figure_mag_force_balances}b.}
\label{Figure_vs_3D}
\end{figure*}

Unlike the QG-MAC crossing $\ell_{\mathrm{MA}}$, the magnetostrophic cross-over $\ell_{\mathrm{MS}}$ does not match with $\ell_{\mathrm{pol}}$ in any of our dynamos models. Therefore, it does not appear to control the dominant scale of convection. It is, however, still possible to relate $\ell_{\mathrm{MS}}$ to the fluid flow. To this end, we decompose the cylindrical radial velocity, $u_{s}$, into its geostrophic ($u_{s}^{g}$) and ageostrophic ($u_{s}^{a}$) components. 3-D renderings of these three velocities are shown in Fig. \ref{Figure_vs_3D}. To construct $u_{s}^{g}$, we average $u_{s}$ along the rotation axis, which yields a flow component that completely satisfies the Proudman-Taylor theorem (Eq. \ref{Proudman_Taylor}). For the integration, we consider the regions inside and outside the inner core tangent cylinder (TC) separately, since the fluid volumes inside the TC have to be averaged independently in the northern and southern hemispheres. Therefore, we compute for
\begin{itemize}[leftmargin=*]
\item[\boldmath$\cdot$] $r_{\mathrm{i}} \leq s \leq r_{\mathrm{o}}$ (i.e. outside TC)
\begin{align}
\bar{u}_{s}^{g} \left(s, \varphi \right) = \frac{1}{2h_{\mathrm{o}}} \int\limits_{\quad -h_{\mathrm{o}}}^{\quad h_{\mathrm{o}}} u_{s} \left( s, \varphi, z \right) \, \mathrm{d}z,
\end{align}
\item[\boldmath$\cdot$] $0 \leq s < r_{\mathrm{i}}$ and $z > 0$ (i.e. inside TC, northern hemisphere)
\begin{align}
\bar{u}_{s}^{g} \left(s, \varphi \right) = \frac{1}{h_{\mathrm{o}} - h_{\mathrm{i}}} \int\limits_{\quad h_{\mathrm{i}}}^{\quad h_{\mathrm{o}}} u_{s} \left( s, \varphi, z \right) \, \mathrm{d}z,
\end{align}
\item[\boldmath$\cdot$] $0 \leq s < r_{\mathrm{i}}$ and $z < 0$ (i.e. inside TC, southern hemisphere)
\begin{align}
\bar{u}_{s}^{g} \left(s, \varphi \right) = \frac{1}{h_{\mathrm{o}} -h_{\mathrm{i}}} \int\limits_{\quad  -h_{\mathrm{o}}}^{\quad - h_{\mathrm{i}}} u_{s} \left( s, \varphi, z \right) \, \mathrm{d}z,
\end{align}
\end{itemize}
where $h_{\mathrm{i}} = \sqrt{r_{\mathrm{i}}^{2}-s^2}$, $h_{\mathrm{o}} = \sqrt{r_{\mathrm{o}}^{2}-s^2}$, and $s$ is the cylindrical radius. 
Subsequently, we expand the $z$-averaged flow $\bar{u}_{s}^{g} \left(s, \varphi \right)$ along the rotation axis back into the initial spherical geometry, which yields the perfectly axially-aligned flow component $u_{s}^{g} \left(r, \theta, \varphi \right)$ visible in Fig.~\ref{Figure_vs_3D}b. The ageostrophic flow component (see Fig.  \ref{Figure_vs_3D}c) is then given by
\begin{align}
u_{s}^{a} = u_{s} - u_{s}^{g}.
\end{align}

\begin{figure*}
\centering
\includegraphics[width=1\textwidth]{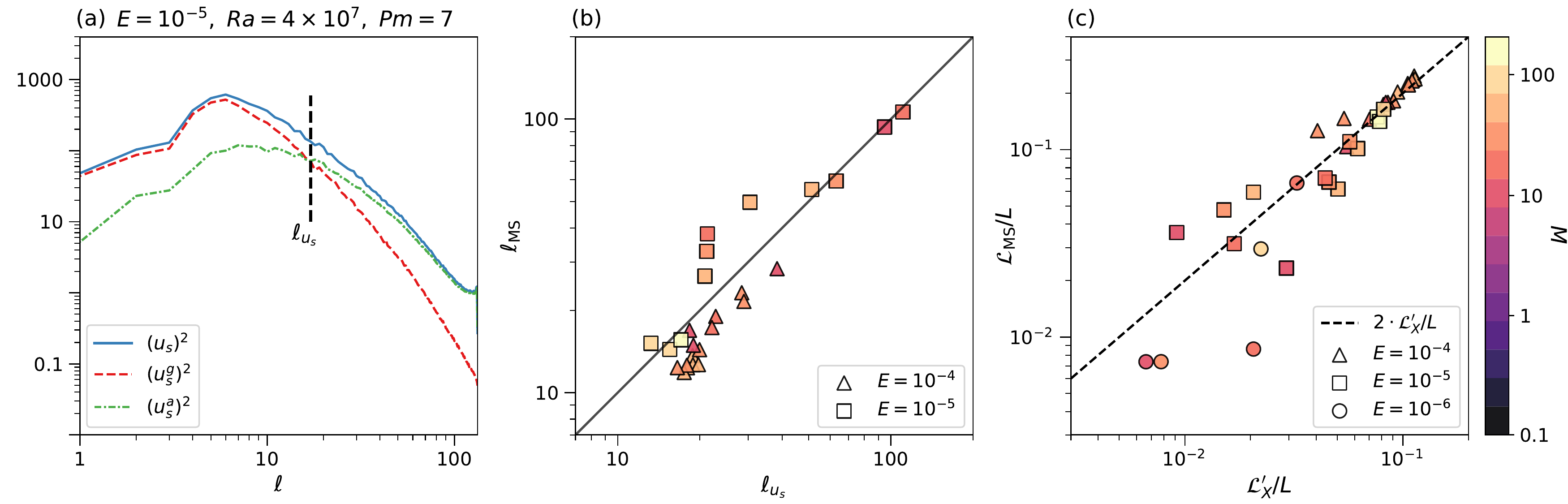}
\caption[]{(a) Example of spectra of the total cylindrical radial velocity $u_{s}$ (solid blue line), its geostrophic component $u_{s}^{g}$ (dashed red line), and its ageostrophic component $u_{s}^{a}$ (dash-dotted green line) for the numerical dynamo at $E=10^{-5}$, $Ra=4\times 10^{7}$ and $Pm=7$. The force balance corresponding to this case is illustrated in Fig.~\ref{Figure_mag_force_balances}b. $\ell_{u_{s}}$ represents the crossing between the spectra of $u_{s}^{g}$ and $u_{s}^{a}$. (b) Comparison of the magnetostrophic crossing $\ell_{\mathrm{MS}}$ defined by the force balance spectra (see Fig. \ref{Figure_mag_force_balances}b) to $\ell_{u_{s}}$. (c) Comparison of $\mathcal{L}_{\mathrm{MS}}/L = \pi/\ell_{\mathrm{MS}}$ with the estimate $\mathcal{L}_{X}^{\prime}/L$ provided by the theoretical scaling  (\ref{same_scale_crossover}). The symbols are coloured with $M$.}
\label{Figure_MS_scale}
\end{figure*}

We construct time-averaged spectra of these three velocities as a function of the degree $\ell$ using spherical harmonic transforms. Figure~\ref{Figure_MS_scale}a illustrates the resulting spectra for the strong-field dynamo whose force balance is shown in Fig.~\ref{Figure_mag_force_balances}b.
We observe that the large scales are almost entirely dominated by the geostrophic flow component, while the small scales are predominantly ageostrophic as expected from the corresponding force balance spectrum. The spherical harmonic degree $\ell_{u_{s}}$ beyond which the flow is mostly ageostrophic does appear to coincide with the magnetostrophic cross-over $\ell_{\mathrm{MS}}$. Repeating this analysis for other cases for which $\ell_{\mathrm{MS}}$ is well-defined, shows that $\ell_{u_{s}}$ and 
$\ell_{\mathrm{MS}}$ generally seem to be in good agreement as shown in Fig.~\ref{Figure_MS_scale}b. This indicates that $\mathcal{L}_{\mathrm{MS}}/L=\pi/\ell_{\mathrm{MS}}$ does indeed represent the length scale beyond which the zeroth-order geostrophy is broken up by the Lorentz force. The large spatial resolutions required for our most extreme simulations restricted this type of analysis to dynamo models with $E \geq 10^{-5}$.

For the comparison of the magnetostrophic cross-over length scales obtained from the force balance spectra, $\mathcal{L}_{\mathrm{MS}}$, to the ones predicted by the Elsasser number scaling suggested by \cite{aurnou_and_king_2017}, $\mathcal{L}_{X}$ (Eq. \ref{magnetostrophic_crossover}), we restrict ourselves to numerical dynamos in which $\ell_{\mathrm{MS}}$ is well-defined and $M \geq 10$. This ensures that the leading-order force balance contains a magnetostrophic range and is not significantly disturbed by inertia and/or viscous forces. $\mathcal{L}_{\mathrm{MS}}$ however does not show any correlation with $\mathcal{L}_{X}$, especially since in our models $\mathcal{L}_{X}$ varies over a much larger range of magnitudes than $\mathcal{L}_{\mathrm{MS}}$. If we slightly modify the scaling by \cite{aurnou_and_king_2017} by assuming similar scales for the fluid flow and the magnetic field, i.e. $\mathcal{L}_{U} \sim \mathcal{L}_{B}$ \citep[e.g.][]{aubert_etal_2017}, instead of different ones and follow the same lines of reasoning as described in Section \ref{elsasser_scaling}, we end up with the following definition for the magnetostrophic cross-over length scale:
\begin{align}
\frac{\mathcal{L}_{X}^{\prime}}{L} = \frac{\Lambda_{\mathrm{t}}}{Rm}.
\label{same_scale_crossover}
\end{align}
The comparison between $\mathcal{L}_{X}^{\prime}/L$ and $\mathcal{L}_{\mathrm{MS}}/L$ is shown in Fig.~\ref{Figure_MS_scale}c. We observe that although the estimate $\mathcal{L}_{X}^{\prime}/L$ does not offer a perfect prediction for $\mathcal{L}_{\mathrm{MS}}/L$, it still gives a reasonable order of magnitude estimate of the magnetostrophic cross-over length scales defined by the force balance spectra.

\subsection{Convective length scales in rotating convection simulations}
\begin{figure*}
\centering
\includegraphics[width=1\textwidth]{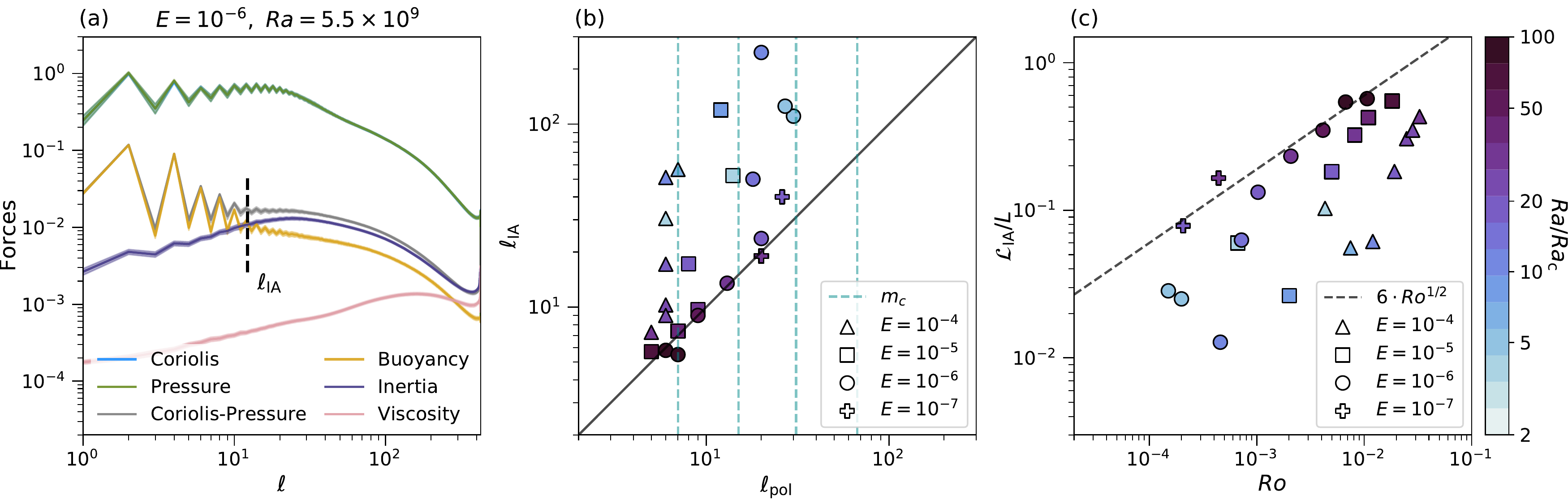}
\caption[]{(a) Example of a time-averaged force balance spectrum of a non-magnetic numerical model with $E = 10^{-6}$, $Ra = 5.5 \times 10^{9}$ ($Ra/Ra_{c} \approx 30$) and $Pr = 1$. The amplitudes of the spherical harmonic contributions of the forces are normalised relative to the peak of the Coriolis force. The shaded regions correspond to one standard deviation in time. $\ell_{\mathrm{IA}}$ is defined as the spherical harmonic degree at which buoyancy and inertia intersect. (b) Comparison of $\ell_{\mathrm{IA}}$ to the peak of the poloidal kinetic energy spectrum $\ell_{\mathrm{pol}}$, which serves as reference for the dominant flow length scale in the rotating convection models. The green vertical dashed lines correspond to the respective azimuthal wavenumber of convection onset $m_{c}$ for $E \in \{ 10^{-4}, 10^{-5}, 10^{-6}, 10^{-7} \}$ from left to right. (c) $\mathcal{L}_{\mathrm{IA}}/L = \pi/\ell_{\mathrm{IA}}$ as a function of $Ro$. The black dashed line corresponds to the theoretical scaling $\mathcal{L}_{\perp}/L \sim Ro^{1/2}$. The symbols in panels (b) and (c) are coloured with the supercriticality $Ra/Ra_{c}$.}
\label{Figure_CIA_scale}
\end{figure*}

In the previous sections, we have shown that the scales at which forces equilibrate in dynamo models correspond to observable convective flow scales. 
In this section, we will now focus on rotating convection models without the influence of a magnetic field, and investigate whether comparable results can be obtained.
Figure~\ref{Figure_CIA_scale}a illustrates the force balance of the non-magnetic counterpart to the dynamo model shown in Fig.~\ref{Figure_mag_force_balances}a.
Similar to the dynamo model, the zeroth-order force balance is geostrophic at all length scales in the non-magnetic case. At the next order, the ageostrophic Coriolis force is now balanced by buoyancy at large scales and inertia towards smaller scales. $\ell_{\mathrm{IA}}$ corresponds to the harmonic degree where the latter two forces are of equal amplitude. Analogously to the magnetic case, we refer to this type of combination of zeroth- and first-order force balance as QG-CIA balance. Viscosity contributes at second order, about one order of magnitude smaller than inertia for the given example. We would like to emphasise, that although both the QG-MAC and the QG-CIA model are geostrophic at zeroth order, the separation between zeroth- and first-order force balance is considerably larger for the latter, indicating a larger degree of geostrophy of the convective flow in the non-magnetic case.

Like for the dynamo models, we again examine whether we can relate the crossing $\ell_{\mathrm{IA}}$ defined by the first-order force balance to the dominant convective flow length scale represented by $\ell_{\mathrm{pol}}$ in spectral space. The comparison of the two length scales, expressed by the associated spherical harmonic degree, is shown in Fig.~\ref{Figure_CIA_scale}b. The symbols corresponding to the individual models are coloured with the supercriticality $Ra/Ra_{c}$ to visualise their proximity to the convection onset. We observe that $\ell_{\mathrm{IA}}$ matches reasonably well with $\ell_{\mathrm{pol}}$ above $Ra/Ra_{c} \gtrsim 25$. In these cases, which are governed by a QG-CIA balance according to the force balance spectra, inertia hence appears to control the flow length scale. For lower supercriticalities, we again observe to a first approximation a clustering of $\ell_{\mathrm{pol}}$ around the critical azimuthal wavenumbers at convection onset for the respective Ekman numbers. This suggests a viscous control of the convective flow in weakly supercritical  models \citep[e.g.][]{gastine_etal_2016,long_etal_2020}.

Given that $\mathcal{L}_{\mathrm{IA}}/L = \pi/\ell_{\mathrm{IA}}$ can be associated with the dominant convective length scale in models that are controlled by a QG-CIA balance, we now want to evaluate whether it follows the predicted Rhines scaling given by relation (\ref{rhines_scaling}). Figure~\ref{Figure_CIA_scale}c shows $\mathcal{L}_{\mathrm{IA}}/L$ as a function of $Ro$. We observe that $\mathcal{L}_{\mathrm{IA}}/L$ starts to approach the $\mathcal{L}_{\perp} / L \sim Ro^{1/2}$ scaling towards the lowest Ekman numbers and largest supercriticalities considered in this study similarly to the results obtained by \citet{gastine_etal_2016} and \citet{guervilly_etal_2019}. However, the length scales for the different Ekman numbers do not collapse on a single curve. This indicates that the analysed models have not fully reached the inertial regime, although they appear to be trending towards it.

\section{Discussion}
\label{discussion}
Several recent studies helped to define the different force balance regimes attained in numerical dynamos by explicitly computing the individual forces \citep{yadav_etal_2016,aubert_etal_2017,aubert_2019,schwaiger_etal_2019}. It, however, still remained an outstanding task to successfully link the force balances to convective flow length scales. Our study suggests that this can be achieved by using the cross-over length scales defined by the spectral representation of force balances introduced by \cite{aubert_etal_2017}. We have found that in dynamos controlled by a QG-MAC balance, the scale at which the Lorentz force and buoyancy balance, $\mathcal{L}_{\mathrm{MA}}$, appears to correspond to the dominant convective scale. This demonstrates the dynamical influence of the Lorentz force in such models. In addition, we have shown  that $\mathcal{L}_{\mathrm{MA}}$ can be reasonably well estimated using the proportionality between large- and small-scale vorticities (Eq. \ref{vorticity_equivalence_nondim}) following the theoretical considerations suggested by \cite{davidson_2013}. Assuming $f_{\mathrm{ohm}}=1$, $Rm=1000$, $\mathcal{L}_{\mathrm{ohm}}=20$ km \citep{aubert_2019} and $L=2260$ km yields $\mathcal{L}_{\mathrm{MA}} \approx 200$ km ($\ell_{\mathrm{MA}} = \pi L / \mathcal{L}_{\mathrm{MA}} \approx 40$) for the Earth's outer core.

The $\mathcal{L}_{\perp} \sim Ro^{1/4}L$ scaling by \cite{davidson_2013}, which additionally assumes rotational independence of the vorticities is not well satisfied by our numerical dynamos. This indicates that this assumption might not be valid in at least the parameter regime covered by our simulations \citep{schrinner_2013}. Possible reasons for this could be that most of the analysed models operate at rather moderate magnetic Reynolds number, $Rm$, and fairly large $Pm$ compared to the Earth's core, where $Rm \sim \mathcal{O} \left( 10^{3} \right)$ and $Pm \sim \mathcal{O} \left( 10^{-6} \right)$. That being said, the exploration of a larger range of $Ro$ might lead to a reassessment of the validity of the scaling law. 
This would however require far more computational resources. Alternatively, the results could possibly be improved by decreasing $Pr$ which would allow a lower $Pm$ to be reached at a given Ekman number.

The second cross-over length scale of the Lorentz force, that is the magnetostrophic cross-over length scale $\mathcal{L}_{\mathrm{MS}}$, can also be retrieved from the convective flow. Our results indicate that it corresponds to the length scale at which the flow dynamics change from predominantly geostrophic (at large scales) to predominantly ageostrophic towards smaller scales. We have shown that a reasonable order of magnitude estimate of $\mathcal{L}_{\mathrm{MS}}$ can be obtained from a modified version of the Elsasser number scaling suggested by \cite{aurnou_and_king_2017} by assuming similar scales for velocity and magnetic fields (Eq. \ref{same_scale_crossover}). Considering $\Lambda_{\mathrm{t}} = 10$ and $Rm = 1000$ \citep{christensen_etal_2010} yields $\mathcal{L}_{\mathrm{MS}} \approx 50$ km ($\ell_{\mathrm{MS}} = \pi L / \mathcal{L}_{\mathrm{MS}} \approx 140$) for the Earth's core.

These results highlight the dynamical influence of the Lorentz force in QG-MAC dynamos, and offer supporting evidences that force balance crossings reflect convective flow length scales. 
We would like to emphasise that our results indicate that it is not necessary for the Lorentz force to reach leading order to be dynamically relevant. 
The dominant scale of convection in QG-MAC dynamos appears to always be controlled by the scale at which buoyancy and Lorentz forces equilibrate in the first-order balance. This is the case independent of whether the zeroth-order balance is geostrophic at all length scales, or subdivided into large-scale geostrophy and small-scale magnetostrophy.

We have obtained comparable results for non-magnetic rotating convection simulations. In models that are controlled by a QG-CIA balance, which is the case at sufficiently high supercriticalities ($Ra/Ra_{c} \gtrsim 25$), the cross-over length scale defined by the first-order force balance, $\mathcal{L}_{\mathrm{IA}}$, again seems to be reflected in the dominant scale of convection.
The comparison of $\mathcal{L}_{\mathrm{IA}}$ to the Rhines scaling $\mathcal{L}_{\perp} \sim Ro^{1/2}L$ suggests that the inviscid regime has not been attained for the Ekman numbers considered in this study. The models at the highest supercriticalities and lowest Ekman numbers, however, appear to approach it.
The fact that the Rhines scaling is not fully met could be attributed to the viscous dissipation in the boundary layers still being significant even at the lowest considered Ekman numbers \citep{gastine_etal_2016}.
The recent study by \cite{guervilly_etal_2019}, in which a quasi-geostrophic approximation of spherical convection is used to reduce the computational costs, suggests that Ekman numbers as low as $E \sim 10^{-9}$ might be required to fully reach the inertial regime. Such extreme parameters are currently still out of reach in 3-D simulations. Given that for the investigated Ekman numbers $\mathcal{L}_{\mathrm{IA}}$ does not fall onto a single curve yet, we assume a prefactor of $10$ for the Rhines scaling, which is slightly larger than what our simulations indicate. Using $\mathcal{L}_{\mathrm{IA}} \sim 10 \, Ro^{1/2} L$ consistent with the results by \cite{guervilly_etal_2019}, yields $\mathcal{L}_{\mathrm{IA}} \approx 20$ km ($\ell_{\mathrm{IA}} = \pi L / \mathcal{L}_{\mathrm{IA}} \approx 350$) for the Earth's outer core in the absence of a magnetic field. For comparison, the viscous scale predicted from Eq. (\ref{viscous_scaling}) would be on the order of twenty meters if $E = 10^{-15}$ is considered. A summary of the theoretical scalings and the predicted flow length scales for the Earth's outer core is given in Table \ref{table_scaling_laws}. Overall, the separation between the different scales is not large, with the exception of the viscous scale, which is about a factor $10^{3}-10^{4}$ smaller than the other scales.

\begin{table}
\caption{Theoretical scaling laws and thereby estimated values for the convective flow length scale in the Earth's outer core. The predictions are obtained by assuming $E=10^{-15}$, $Ro=10^{-6}$, $Rm=1000$, $\Lambda=10$, $\mathcal{L}_{\mathrm{ohm}} = 20$ km and $L=2260$ km.}
\centering
\begin{tabular}{cccc}
\hline
Force balance & Scaling law & Predicted value \\ \hline
QG-VAC	& $\mathcal{L}_{\perp} \sim E^{1/3} L$ & $\sim 20$ m \\[0.5mm]
QG-CIA	& $\mathcal{L}_{\perp} \sim Ro^{1/2} L$ & $\sim 20$ km \\[0.5mm]
QG-MAC	& $\mathcal{L}_{\perp} \sim Rm \ \mathcal{L}_{\mathrm{ohm}}^{2} / L$ & $\sim 200$ km \\[0.5mm]
QG-MAC	& $\mathcal{L}_{\perp} \sim Ro^{1/4} L$ & $\sim 100$ km \\[0.5mm]
QG-MS	& $\mathcal{L}_{X}^{\prime} \sim \Lambda L / Rm$ & $\sim 50$ km \\ \hline
\end{tabular}
\label{table_scaling_laws}
\end{table}

\section{Conclusions}
\label{conclusions}
Our main findings can be summarised by the following points:

(\textit{i}) Length-scale-dependent analysis of the forces shows that in QG-MAC dynamos, the scale at which buoyancy and Lorentz forces are of equal magnitude is reflected in the dominant scale of convection. Similar results hold for non-magnetic rotating convection simulations that are controlled by a QG-CIA balance. In these models, the scale at which buoyancy and inertia equilibrate can be associated to the primary convective scale.

(\textit{ii}) In most QG-MAC dynamos, the dominant force balance is divided into large-scale geostrophy and small-scale magnetostrophy.  The analysis of the axial invariance of the flow reveals that the length scale which marks this transition can be retrieved from the convective pattern. In agreement with the prevailing force balance, variations along the axis of rotation are close to invariant along the rotation axis on large scales, while this is much less the case on smaller scales.

(\textit{iii}) Assuming that the Earth's dynamo is controlled by a QG-MAC balance, analysing the scaling behaviour of the two characteristic length scales found in geodynamo models suggests that the dominant flow length scale in the Earth's core is about $200 \, \mathrm{km}$, while magnetostrophic effects are deferred to scales smaller than $50 \, \mathrm{km}$.

Future contributions to the discussion on convective flow length scales in numerical dynamo models could involve the analysis of the $z$-averaged flow due to the prevailing (large-scale) geostrophy, as has been done for the $3$-D models of non-magnetic rotating convection in \citet{guervilly_etal_2019}. This would allow  access to the integral flow length scale $\mathcal{L}_{\perp}$ perpendicular to the rotation axis. A similar strategy could potentially also be applied to the computation of the force balance, which might provide additional insights on how to relate force balances to convective flow length scales.

\begin{acknowledgments}
We would like to thank two anonymous reviewers for their constructive comments. 
We acknowledge support from the Fondation Del Duca of Institut de France (JA, 2017 Research Grant). Numerical computations were performed at S-CAPAD, IPGP and using HPC resources from GENCI-CINES and GENCI-TGCC (Grants A0040402122 and A0060402122). This project has received funding from the European Union's Horizon 2020 research and innovation programme under the Marie Sk\l{}odowska-Curie grant agreement No~665850. All figures have been generated using either \texttt{matplotlib} \citep{hunter_2007} or \texttt{paraview}.
\end{acknowledgments}

\bibliographystyle{abbrvnat}
\interlinepenalty=10000

\appendix
\begin{table*}
\caption{Summary of the relevant parameters of the non-magnetic rotating convection simulations considered in the present study. All models have been computed with $Pr = 1$ and $r_{\mathrm{i}}/r_{\mathrm{o}}=0.35$. $Nu$ represents the Nusselt number.}
\centering
\begin{tabular}{cccccccc}
\toprule
{} &                  $Ra$ &     $Nu$ &                  $Ro$ &       $Re$ & $\ell_{\mathrm{pol}}$ & $\ell_{\mathrm{IA}}$ &  $\left( N_{r},\ell_{\mathrm{max}} \right)$ \\
\midrule
\addlinespace[0.1cm]
\multicolumn{8}{c}{$E=10^{-4}$} \\
\addlinespace[0.1cm]
$1$  &  $3.000\times 10^{6}$ &   $1.51$ &  $4.31\times 10^{-3}$ &    $43.15$ &                 $6$ &              $31.18$ &   $\left( 41,85 \right)$ \\
$2$  &  $4.830\times 10^{6}$ &   $2.13$ &  $7.46\times 10^{-3}$ &    $74.56$ &                 $7$ &              $55.57$ &   $\left( 41,85 \right)$ \\
$3$  &  $7.500\times 10^{6}$ &   $3.02$ &  $1.20\times 10^{-2}$ &   $120.12$ &                 $6$ &              $52.66$ &   $\left( 41,85 \right)$ \\
$4$  &  $1.125\times 10^{7}$ &   $4.67$ &  $1.89\times 10^{-2}$ &   $189.28$ &                 $6$ &              $23.43$ &   $\left( 41,85 \right)$ \\
$5$  &  $1.500\times 10^{7}$ &   $5.87$ &  $2.48\times 10^{-2}$ &   $247.81$ &                 $6$ &               $9.58$ &   $\left( 41,85 \right)$ \\
$6$  &  $1.750\times 10^{7}$ &   $6.51$ &  $2.82\times 10^{-2}$ &   $282.45$ &                 $6$ &               $8.26$ &   $\left( 65,128 \right)$ \\
$7$  &  $2.100\times 10^{7}$ &   $7.27$ &  $3.27\times 10^{-2}$ &   $327.17$ &                 $5$ &               $6.37$ &   $\left( 65,133 \right)$ \\
\addlinespace[0.1cm]
\multicolumn{8}{c}{$E=10^{-5}$} \\
\addlinespace[0.1cm]
$8$ &  $4.000\times 10^{7}$ &   $1.39$ &  $6.66\times 10^{-4}$ &    $66.63$ &                $14$ &              $51.31$ &   $\left( 97,170 \right)$ \\
$9$ &  $1.000\times 10^{8}$ &   $2.44$ &  $2.01\times 10^{-3}$ &   $201.38$ &                $12$ &              $120.4$ &   $\left( 97,213 \right)$ \\
$10$ &  $2.000\times 10^{8}$ &   $5.41$ &  $4.99\times 10^{-3}$ &   $499.27$ &                 $8$ &               $14.99$ &   $\left( 97,256 \right)$ \\
$11$ &  $3.000\times 10^{8}$ &   $9.01$ &  $8.15\times 10^{-3}$ &   $815.16$ &                 $9$ &               $8.66$ &  $\left( 121,288 \right)$ \\
$12$ &  $4.000\times 10^{8}$ &  $11.95$ &  $1.09\times 10^{-2}$ &  $1092.02$ &                 $7$ &               $6.33$ &  $\left( 121,288 \right)$ \\
$13$ &  $7.000\times 10^{8}$ &  $18.13$ &  $1.82\times 10^{-2}$ &  $1824.76$ &                 $5$ &               $4.69$ &  $\left( 161,426 \right)$ \\
\addlinespace[0.1cm]
\multicolumn{8}{c}{$E=10^{-6}$} \\
\addlinespace[0.1cm]
$14$ &  $8.000\times 10^{8}$ &   $1.42$ &  $1.51\times 10^{-4}$ &   $150.56$ &                $30$ &             $110.64$ &  $\left( 129,256 \right)$ \\
$15$ &  $1.000\times 10^{9}$ &   $1.55$ &  $2.00\times 10^{-4}$ &   $199.62$ &                $27$ &             $125.53$ &  $\left( 129,256 \right)$ \\
$16$ &  $2.000\times 10^{9}$ &   $2.38$ &  $4.59\times 10^{-4}$ &   $458.87$ &                $20$ &             $244.43$ &  $\left( 129,341 \right)$ \\
$17$ &  $2.800\times 10^{9}$ &   $3.42$ &  $7.21\times 10^{-4}$ &   $721.33$ &                $18$ &             $43.13$ &  $\left( 161,426 \right)$ \\
$18$ &  $3.500\times 10^{9}$ &   $4.74$ &  $1.03\times 10^{-3}$ &  $1028.67$ &                $20$ &              $22.49$ &  $\left( 181,426 \right)$ \\
$19$ &  $5.500\times 10^{9}$ &   $9.42$ &  $2.05\times 10^{-3}$ &  $2045.89$ &                $12$ &              $12.25$ &  $\left( 201,426 \right)$ \\
$20$ &  $1.000\times 10^{10}$ & $19.68$ &  $4.13\times 10^{-3}$ &  $4125.75$ &                 $9$ &       $6.50$  &  $\left( 321,512 \right)$ \\
$21$ &  $1.600\times 10^{10}$ & $30.78$ &  $6.71\times 10^{-3}$ &  $6714.67$ &                 $6$ &       $4.78$  &  $\left( 385,597 \right)$ \\
$22$ &  $2.660\times 10^{10}$ & $46.46$ &  $1.06\times 10^{-2}$ & $10613.04$ &                 $5$ &       $4.46$  &  $\left( 513,853 \right)$ \\
\addlinespace[0.1cm]
\multicolumn{8}{c}{$E=10^{-7}$} \\
\addlinespace[0.1cm]
$23$ &  $6.500\times 10^{10}$ &  $4.15$ &  $2.07\times 10^{-4}$ &  $2074.51$ &                $26$ &     $34.83$ &  $\left( 433,682 \right)$ \\
$24$ &  $1.000\times 10^{11}$ &  $7.92$ &  $4.42\times 10^{-4}$ &  $4420.00$ &                $20$ &     $19.00$ &  $\left( 513,682 \right)$ \\
\bottomrule
\end{tabular}
\label{appendix_hydro_table}
\end{table*}

\label{lastpage}

\end{document}